\documentclass[11pt]{article}

\usepackage[margin=0.6in]{geometry}
\usepackage{setspace}
\usepackage{amsmath,amssymb,amsthm,bm,mathrsfs,bbold}
\usepackage{accents}
\usepackage{prodint}

\usepackage{graphicx}
\usepackage{float}
\usepackage{booktabs}
\usepackage{multirow}
\usepackage{array}
\usepackage{adjustbox}

\usepackage[dvipsnames]{xcolor}

\usepackage{caption}
\usepackage{subcaption}
\usepackage[
    colorlinks=true,
    linkcolor=MidnightBlue,
    citecolor=MidnightBlue,
    urlcolor=MidnightBlue
]{hyperref}

\usepackage[authoryear]{natbib}
\setcitestyle{authoryear,open={(},close={)}}

\usepackage{enumerate}
\usepackage{enumitem}
\usepackage{moreverb}
\usepackage{hyphenat}
\usepackage{url}
\usepackage{tikz}
\usepackage{tcolorbox}
\usepackage{wasysym}

\usepackage[small]{titlesec}

\titleformat{\subsubsection}[runin]
{\normalfont\bfseries}
{\thesubsubsection}
{1em}
{}

\newtheorem{assumption}{Assumption}

\newcommand{\redd}[1]{\textcolor{red!80}{#1}}

\newcommand{\EE}{\mathbb{E}}

\newcommand{\R}{\mathbb{R}}

\newcommand{\KK}{K} 
\newcommand{\1}{\mathbb{1}}

\newcommand{\eps}{\varepsilon}

\newcommand\independent{%
  \protect\mathpalette{\protect\independenT}{\perp}
}
\def\independenT#1#2{%
  \mathrel{\rlap{$#1#2$}\mkern2mu{#1#2}}
}

\title{Your Title Here}

\author{
First Author\\
Institution
\and
Second Author\\
Institution
}

\date{\today}

\begin{document}

\title{ Estimation of treatment effects in presence of differential
  use of post-randomization concomitant medication with time-to-event outcomes
  %
}

\author{ \small
Helene C. W. Rytgaard$^{1}$,
Edwin Fong$^{2}$,
Jens M. Tarp$^{3}$,
Thomas A. Gerds$^{1}$,
Mark J. van der Laan$^{4}$,
Henrik Ravn$^{3}$
\\[1em]
\small $^{1}$Section of Biostatistics, University of Copenhagen, Copenhagen, Denmark
\\
\small $^{2}$Department of Statistics and Actuarial Science, The University of Hong Kong, Pokfulam, Hong Kong, China
\\
\small $^{3}$Novo Nordisk A/S, Søborg, Denmark
\\
\small $^{4}$Division of Biostatistics, School of Public Health, and Department of Statistics,
University of California, California, USA
\\[1em]
\small Corresponding author:
Helene C. W. Rytgaard
\\
\small Øster Farimagsgade 5, 1014 Copenhagen, Denmark
\\
\small \texttt{hely@sund.ku.dk}
}

\begin{singlespace}
\maketitle
\end{singlespace}

\abstract{

  In placebo-controlled randomized trials, the post-randomization use
  of concomitant medications may be higher in the placebo arm than in
  the treatment arm. This may dilute the full benefits of the
  randomized drug as estimated by the intention-to-treat
  analysis. We focus on cardiovascular outcomes trials in type-2
  diabetes patients of glucose-lowering treatments where patients in
  the placebo arm are more likely to add other glucose-lowering agents
  with established cardio-protective properties. As a supplement to
  the intention-to-treat analysis, we propose a class of estimands
  within a causal framework that isolates the specific impact of the
  treatment being studied from that of concomitant treatment
  use. These estimands are defined under time-dependent treatment
  interventions to balance exposure to additional medications across
  intervention arms. We advocate for specific stochastic interventions
  to achieve this balance while minimizing positivity violations,
  which arise when certain treatment combinations or characteristics
  are not sufficiently represented in the data. We employ targeted
  minimum loss-based estimation (TMLE) to optimize the estimation
  procedure for our estimands while allowing for ﬂexible adjustments
  for time-dependent covariates from follow-up visits.  Finally, we
  demonstrate the application of the methods through a simulation
  study and a real-world example from the LEADER cardiovascular
  outcomes trial, which assessed cardiovascular risk for liraglutide
  versus placebo.

}



\footnotetext{\textbf{Abbreviations:} TMLE, targeted minimum
   loss-based estimation; ITT, intention-to-treat; CVOT, cardiovascular
   outcomes trial; MACE, major adverse cardiovascular event; GLP1,
   glucagon-like peptide-1}

\newpage
 
\section{Introduction}\label{sec1}

In randomized controlled trials (RCTs), participants are enrolled and
randomized to either a treatment or placebo arm. Randomization ensures
that the treatment assignment is unconfounded and that use of
concomitant treatments is balanced at baseline. The primary
statistical analysis typically follows the intention-to-treat (ITT)
principle, where subjects are analyzed according to their original
randomization, regardless of adherence to the assigned treatment or
use of additional medications. This is often referred to as a
\textit{treatment policy strategy}\cite{international2019addendum},
where intercurrent events, such as additional medication use,
treatment discontinuation, or dose changes, are considered irrelevant
for defining the treatment effect.

In placebo-controlled cardiovascular outcomes trials of
glucose-lowering treatments, the increasing availability of already
approved glucose-lowering agents with cardio-protective properties
introduces a significant challenge to the analysis of trial results
\cite{bethel2020exploring,mcguire2022transitioning}. These concomitant
medications, when used by participants in either arm of the trial, can
dilute or mask the full benefits of the randomized treatment
.
Particularly, 
unbalanced usage of concomitant medications across treatment arms
complicates the isolation of the randomized treatment's full
impact. This can diminish the expected differences between the
intervention and placebo arms, potentially necessitating larger sample
sizes to detect effects. Consequently, conducting an adequately
powered randomized controlled trial (RCT) in this context may even be
unfeasible.
These constraints highlight the need for more sophisticated
statistical estimation techniques, such as those based on time-varying
treatment interventions and estimand frameworks, to account for the
presence of concomitant medications and allow for estimation of the
effect of the randomized treatment while controlling for unbalanced
usage across treatment arms
.

In this work, we present statistical methods for conceptualizing
hypothetical interventions and applying targeted learning estimation
to address the challenge of isolating the specific impact of the
treatment being studied from that of concomitant treatment use. In our
analysis, we use a precise definition and description of a treatment
effect linked to a specific clinical question, which can be captured
through the choice of an estimand that summarizes the distributional
features of outcomes had subjects been followed under different
treatment conditions \cite{hernan2020causal}.
The estimand defined in accordance with the
intention-to-treat principle reflects a treatment policy strategy,
wherein intercurrent events, such as the use of additional medications
or deviations from the randomized treatment, are considered part of
the treatments being compared.
Other treatment effects can be derived under hypothetical strategies
that correspond to different conditions than those observed in the
trial \cite{petersen2014causal,hernan2016using}. Particularly, while
ethical concerns about withholding proven therapies from the control
(placebo) group may limit the extent to which pure placebo-controlled
comparisons can be conducted, the effect when hypothetically
preventing concomitant treatment use can, in principle, be defined and
estimated with appropriate statistical methods
\cite{hernan2020causal}.
%

\begin{figure}[ht]
  \vspace{-5mm}
  \centering \includegraphics[width=0.75\textwidth,angle=0]
  {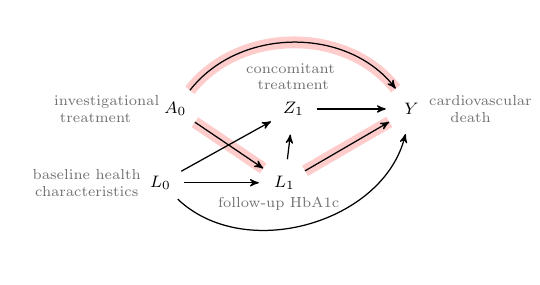}
  \vspace{-12mm}
  \caption{Simplified causal diagram to show the targeted direct
    effect of the randomized treatment.  }
  \label{fig:intro:simple:dag}
\end{figure}

Figure \ref{fig:intro:simple:dag} illustrates, with a simplified
causal diagram including only two time-points and few key variables,
the direct pathway and a potential diluting pathway from the
randomized glucose-lowering treatment to the cardiovascular
outcome.  To reflect the RCT situation, there is no arrow from
baseline health characteristics to randomized treatment due to
randomization. Furthermore, since the randomized treatment assignment
is blinded, and neither the trial participant nor their doctor knows
whether the trial participant received the active treatment or
placebo, there is no arrow from randomized treatment to follow-up
concomitant treatment decisions. Follow-up concomitant treatment
decisions are instead based on patient health, and the diagram depicts
a simplified version of reality where the decision is made based on
baseline characteristics and the follow-up HbA1c value. This
simplification is not an unreasonable reflection of clinical practice,
as clinical guidelines generally recommend initiating additional
glucose-lowering treatment when the HbA1c value reaches a certain
level, representing poor glycemic control. Since individuals in the
placebo group, lacking the protective effects of the randomized
treatment, are much more likely to experience poor glycemic control,
they are consequently more likely to initiate concomitant medication.
The targeted direct effect of the randomized treatment on the
cardiovascular outcome corresponds to the highlighted pathways from
randomized treatment to outcome in the figure.

In mediation analysis
\cite{robins1992identifiability,pearl2001direct}, one distinguishes
between the \emph{controlled direct effect}, which is the effect that
the randomized treatment has on the outcome while the exposure to
concomitant treatment is fixed to some pre-specified value, and the
\emph{natural direct effect} which is the effect of the randomized
treatment on the outcome while the exposure to concomitant treatment
takes its value as if the randomized treatment were absent. While
controlled direct effects may readily be defined via hypothetical
static interventions enforcing, for example, no exposure to
concomitant treatment at any time (corresponding to censoring
individuals at the time of concomitant treatment initiation), such
interventions typically do not have support in real data settings,
leading to positivity violations that can significantly compromise the
statistical estimation procedure \citep{petersen2012diagnosing}.
To mitigate such positivity violations, we propose in this work
particular stochastic interventions that, like the static
interventions, balance exposure to concomitant treatment across the
active treatment and the placebo arm, but closer resembles what was
actually observed in the observed trial. The corresponding effects can
be interpreted like natural direct effects related to those proposed
in the mediation literature
\cite{didelez2006direct,van2008direct,vanderweele2017mediation,zheng2017longitudinal,vansteelandt2019mediation,michiels2021novel},
with the distinction that our interventional distributions are defined
conditional on baseline covariates, longitudinal survival status and
follow-up information on concomitant treatment history, but not
post-baseline covariates. As such, this aligns with the random
intervention approach with marginal mediator distributions of
VanderWeele and Tchetgen Tchetgen
(2017)\cite{vanderweele2017mediation}, but differs in that we
condition on individual survival status. Referring again to Figure
\ref{fig:intro:simple:dag}, our approach isolates the highlighted
direct pathway, particularly from the indirect pathway leading from
the randomized treatment to follow-up covariates (glycemic control in
the figure) and then to follow-up decisions regarding concomitant
treatment usage. The figure also illustrates that while interventions
conditioned on post-baseline covariates are inherently more realistic
than those defined without consideration of these covariates, they do
not yield the targeted interpretation. In fact, if the causal diagram
accurately reflects reality, these interventions would not differ from
the intention-to-treat effect.

We emphasize the general importance of appropriately addressing
time-dependent confounding whenever the target of estimation involves
hypothetically altering the distribution of concomitant medication,
whether this involves simple censoring-type interventions or more
general stochastic interventions. As highlighted, initiation of
concomitant medication is typically driven by elevated HbA1c, which is
also a strong predictor of cardiovascular death, introducing
time-varying confounding by indication. Without proper adjustment for
HbA1c, a naive approach that for example treats concomitant treatment
initiation as a censoring event when fitting a Cox regression model
with only baseline treatment would tend to remove higher-risk
individuals from the placebo arm, biasing results in favor of placebo.
Similarly, any model specification conditional on HbA1c would block
previous treatment effects on the outcome going through better control
of HbA1c \citep{robins1986new} and thus similarly bias the results in
favor of placebo.
Inverse probability weighting, such as used by Bethel et
al. (2020)\cite{bethel2020exploring}, on the other hand, offers an
approach to estimate the effect of the randomized treatment measured
as a hazard ratio when censoring for concomitant treatment initiation
which does allow for properly controlling for the reasons to start
concomitant treatment. However, this approach presents two key
limitations, in addition to the highlighted positivity problems with
such controlled (rather than natural) direct effects:
\begin{enumerate}
\item Hazards ratios are limited in their reflection of treatment
  effects, suffering from inherent causal interpretation problems by
  mixing treatment effects with selection
  bias\cite{hernan2010hazards,martinussen2020subtleties}. In contrast,
  we focus on absolute risks as the main target for estimation of
  treatment effects, which avoid these pitfalls and offer a clearer
  measure of the effect of treatments on cardiovascular risks.
\item Correct adjustment for time-dependent confounding requires
  correct model specification or the use of data-adaptive machine
  learning methods, and this is not possible in combination with
  inverse probability weighting while still providing valid
  statistical inference.
\end{enumerate}

%
%
To optimize the estimation procedure for the target estimands defined
as contrasts of intervention\hyp{}specific absolute risks, and allow
for data-adaptive adjustment for time-dependent confounding
\cite{robins1986new,hernan2020causal}, we utilize longitudinal
targeted minimum loss-based estimation (TMLE)
\cite{van2006targeted,van2011targeted,stitelman2011rcts,zheng2017longitudinal,ltmleRpackage}. In
this estimation framework, semiparametric efficient and double robust
estimation is combined with flexible machine learning based estimation
and asymptotic statistical inference based on the efficient influence
curve\cite{bang2005doubly}.  We outline the key steps of the
estimation procedure for longitudinal TMLE, which are necessary for
users of the methods, in our setting with stochastic interventions,
time-dependent confounding, terminal outcome events, and competing
risks, and we describe how super learning\cite{van2007super,polley2011super} can
be utilized to flexibly learn how follow-up treatment decisions and
outcome risks depend on previous health measurements and treatment
usage.
For illustration, we consider the LEADER cardiovascular outcomes trial
which included 9,340 patients with type-II diabetes and high
cardiovascular risk, randomized to either treatment with liraglutide
(a once-daily injected glucagon-like peptide-1, abbreviated GLP1) or
placebo. The primary analysis was an intention-to-treat analysis based
a Cox regression model with treatment as the only variable and
estimated a hazard ratio of 0.87 with a 95\% confidence interval from
0.78 to 0.97 \cite{marso2016liraglutide}.  We use this study to
illustrate our methods, addressing the potential issues arising
particularly with respect to use of insulin; while the cardiovascular
benefits of insulin may not severely influence the concluded treatment
effect of GLP1, the methodological issues we tackle are a growing
problem in randomized controlled trials
\cite{mcguire2022transitioning}. We also present a brief simulation
study designed to mimic the data analysis setting, illustrating how
such data could have arisen and highlighting differences in positivity
violations under different hypothetical interventions.

\section{Setting and notation}
\label{sec:observed:data:structure}

We consider an observed data setting as follows, with subject-specific
information collected in a vector \(O\) consisting of information
obtained across the trial initiation (time \(t_0\)) and
\(k=1,\ldots, K\) follow-up visits in the trial:
\begin{align}
  O = (L_0, Z_0, A_0, Y_1, D_1, C_1, L_1, Z_1, A_1, \ldots, 
  Y_{K-1},D_{K-1}, C_{K-1}, L_{K-1}, Z_{K-1}, A_{K-1}, Y_{K}) . 
\label{eq:unit:O}
\end{align}
In this notation, \(L_0 \in \R^d\) denotes baseline covariates,
\(Z_0\in \lbrace 0,1\rbrace\) denotes the indicator of use of
concomitant medication at the time of randomization, and
\(A_0\in\lbrace 0,1\rbrace\) is an indicator of randomized
treatment arm assigned to (active treatment versus placebo). Follow-up
treatment information is collected in variables
\(A_k\in \lbrace 0,1\rbrace\) and \(Z_k\in \lbrace 0,1\rbrace\)
defined as indicators of exposure to the randomized treatment and
concomitant treatment, respectively, at visit \(k\), so that \(A_k\)
particularly captures adherence or non-adherence to initial
randomization arm. Time-dependent (post-randomization) covariates are
collected in the vector \(L_k\in\R^{d'}\). Finally,
\(Y_k\in \lbrace 0,1\rbrace\), \(D_k\in \lbrace 0,1\rbrace\) and
\(C_k\in \lbrace 0,1\rbrace\) indicate the primary endpoint
(cardiovascular death), death (of other causes) and censoring status,
respectively, at visit \(k\). We use \(t_k\) to denote the time in
months between time of randomization (\(t_0\)) and visit number \(k\).
With \(\tilde{T} = \min (T,C)\) denoting the minimum of the time \(T\)
to event and the time \(C\) to censoring and
\(\tilde{\Delta} = \1\lbrace T\le C\rbrace \Delta \in \lbrace
0,1,2\rbrace\) an indicator of event (\(\tilde{\Delta}=\Delta=1\)),
competing death (\(\tilde{\Delta}=\Delta=2\)) or censoring
(\(\tilde{\Delta}=0\)), we have that
\(Y_k = \1\lbrace \tilde{T} \le t_k, \tilde{\Delta} =1 \rbrace\),
\(D_k = \1\lbrace \tilde{T} \le t_k, \tilde{\Delta} =2 \rbrace\) and
\(C_k = \1\lbrace \tilde{T} \le t_k, \tilde{\Delta} =0 \rbrace\).  The
trial population corresponds to a sample
\(\lbrace O_i \rbrace_{i=1}^n\) of independent and identically
distributed observations defined in Equation \eqref{eq:unit:O}. The
data-generating distribution is denoted \(P_0\) and assumed to belong
to a statistical model \(\mathcal{M}\).

Throughout, we assume the following temporal ordering among the
observed variables across time-points \(t_k\):
\begin{align}
\begin{Bmatrix}
    L_0 \\
    Z_{0}
\end{Bmatrix}  \rightarrow  
  A_{0} 
  \rightarrow \cdots \rightarrow L_{k-1} \rightarrow
    \begin{Bmatrix}
    A_{k-1} \\
    Z_{k-1}
  \end{Bmatrix} \rightarrow \begin{Bmatrix}
    Y_{k} \\
    D_{k} \\ C_{k}
  \end{Bmatrix}\rightarrow L_{k} \rightarrow
  \begin{Bmatrix}
    A_k \\
    Z_k
  \end{Bmatrix}\rightarrow \begin{Bmatrix}
    Y_{k+1} \\
    D_{k+1} \\ C_{k+1}
  \end{Bmatrix} \rightarrow \cdots;
  \label{eq:ordering}
\end{align}
particularly, we assume that the covariates \(L_k\) measured at time
\(t_k\) may depend on all previous treatment decisions
\(A_0, Z_0, \ldots, A_{k-1},Z_{k-1}\), but not the treatment decisions
\(A_k\) and \(Z_k\) \textit{at} that time, whereas the treatment
decisions \(A_k\) and \(Z_k\) at time \(t_k\) may be influenced by the
entire history of covariates \(L_0 \ldots,L_{k}\) up to time \(t_k\).
For a given variable \(X_k\), we denote by \(\mathrm{Pa}(X_k)\) the
parent set of variables occurring before \(X_k\) in the observed
variable ordering. We further denote by
\(\bar{A}_k = (A_0, A_1, \ldots A_k)\) the history of randomized
treatment exposure up to time \(t_k\), and use the notation
\(\bar{Z}_k\) and \(\bar{L}_k\) similarly. Note that an overview of
key notation can also be found in the Supplementary Material.

\subsection{
  Observed data distribution}

We assume that the observed data distribution \(P_0\) belongs to a
statistical model \(\mathcal{M}\) consisting of distributions \(P\)
factorized according to the observed variable ordering. We here
introduce notation to express this factorization. Let
\(q_{L_k}( \ell \mid \mathrm{Pa}(L_k) )\) denote the conditional
density of covariates \(L_k\) with respect to a dominating measure
\(\nu_{L_k}\). Let
\(q_{Y_k}( y \mid \mathrm{Pa}(Y_k)) = P(Y_k= y \mid
\mathrm{Pa}(Y_k))\) denote the conditional distribution of primary
outcome status \(Y_k\), and let
\(q_{D_k}( d \mid \mathrm{Pa}(D_k)) = P(D_k= d \mid
\mathrm{Pa}(D_k))\) denote the conditional distribution of death
status \(D_k\).  For the treatment variables, let
\(g_{A_k} ( a \mid \mathrm{Pa}(A_k)) = P(A_k = a \mid
\mathrm{Pa}(A_k)) \) denote the conditional distribution of
randomized treatment status \(A_k\), and let
\(g_{Z_k} ( z \mid \mathrm{Pa}(Z_k)) = P(Z_k = z \mid
\mathrm{Pa}(Z_k)) \) denote the conditional distribution of
concomitant treatment status \(Z_k\). Finally, let
\(g_{C_k}( c \mid \mathrm{Pa}(C_k)) = P(C_k= c \mid
\mathrm{Pa}(C_k))\) denote the conditional distribution of censoring
status \(C_k\). The density \(p\) of the distribution \(P\) factorizes
as follows:
\begin{align}
  \begin{split}
    p (O) = \prod_{l=0}^{K-1}
    g_{A_l}(A_l \, \vert \, \mathrm{Pa}(A_l))
    g_{Z_l}(Z_l \, \vert \, \mathrm{Pa}(Z_l)) 
    \prod_{l=0}^{K-1}
    q_{L_{l}} ( L_{l} \mid  \mathrm{Pa}(L_{l})) 
    \\
    \prod_{l=1}^{K-1} \big( g_{C_l}(1 \, \vert \, \mathrm{Pa}(C_l))
    \big)^{C_l} \big( g_{C_l}(0 \, \vert \,
    \mathrm{Pa}(C_l))\big)^{1-C_l}
    \\
    \prod_{l=1}^{K-1} \big( q_{D_l}(1 \, \vert \, \mathrm{Pa}(D_l))
    \big)^{D_l} \big( q_{D_l}(0 \, \vert \,
    \mathrm{Pa}(D_l))\big)^{1-D_l}
    \\
    \prod_{l=1}^{K} \big( q_{Y_l}(1 \, \vert \, \mathrm{Pa}(Y_l))
    \big)^{Y_l} \big( q_{Y_l}(0 \, \vert \, \mathrm{Pa}(Y_l))\big)^{1-
    Y_l} .\end{split}
  \label{eq:like}
\end{align}
{Note that we have used the notation `\(q\)' for some factors and the
  notation `\(g\)' for other factors to reflect what will later be
  subject to hypothetical interventions.  We decompose any
  \(P\) into two parts accordingly, referring to these
  parts as the interventional part (\(g\)) and the non-interventional
  part (\(q\)), and write
  \begin{align}
    {p}= {p_{q,g}} = \prod_{l = 0}^{K} q_{l} \, g_{l},
    \label{eq:P:QG}
  \end{align}
  where \(g_{l}\) and \(q_{l}\) denote products at time \(t_l\) of
  conditional densities of non-interventional nodes and interventional
  notes, respectively. The statistical \(\mathcal{M}\) can be represented as
  \(\mathcal{M}= \lbrace {p}= {p_{q,g}} \, : \, q\in\mathcal{Q},
  q\in\mathcal{G}\rbrace\), where \(\mathcal{Q}\) is parameter set of
  possible values for \(q\) and \(\mathcal{G}\) is parameter set of
  possible values for \(g\). We generally assume that \(\mathcal{Q}\)
  and \(\mathcal{G}\) are subsets of broad function spaces, and
  particularly that \(\mathcal{M}\) is a proper nonparametric
  model. However, it is worth mentioning that we do have some knowledge
  on \(\mathcal{G}\), specifically that the randomized treatment is
  randomized at baseline. Nevertheless, this assumption, as well as any
  non-testable assumptions on \(\mathcal{G}\), does not impact the
  statistical estimation problem \cite{van2011targeted}. 
}

\section{Estimands isolating the direct impact of randomized
  treatment}
\label{sec:causal:questions}


Recall that our primary objective is to isolate the direct impact of
the randomized treatment, distinct from indirect effects it may
have through imbalance in use of concomitant outcome-protective
medications. In this section, we introduce hypothetical interventions
on the distributions governing treatment decisions \(A_k\) and \(Z_k\)
over time, to emulate hypothetical scenarios where concomitant
treatment exposure is balanced between the treatment and placebo arms,
and where trial participants fully adhere to their assigned
randomization arms.
Section \ref{sec:hypothetical:interventions} defines these
interventions, 
while Section \ref{sec:target:estimand} defines corresponding target
causal estimand(s) as contrasts of intervention\hyp{}specific absolute
risks.  
The identifiability assumptions required for these estimands,
depending on the intervention choice, are discussed in Section
\ref{sec:identifiability}.

\subsection{Time-dependent treatment interventions}
\label{sec:hypothetical:interventions}



In the real trial, the trial participant (while alive and under
observation, i.e., \(\bar{Y}_k=\bar{D}_k=\bar{C}_k=0\)), and/or their
doctor consider baseline characteristics, \(L_0\), randomized
treatment adherence and concomitant treatment usage so far,
\(\bar{A}_{k-1}, \bar{Z}_{k-1}\), along with changes in health
measurements, \(\bar{L}_{k}\), and makes a decision to continue or
change treatment according to the observed-data distributions
\(g_{A_k} ( a \mid \bar{D}_k, \bar{Y}_k, \bar{C}_k,
\bar{L}_{k},\bar{A}_{k-1}, \bar{Z}_{k-1}, L_0) = P(A_k = a \mid
\bar{D}_k, \bar{Y}_k, \bar{C}_k, \bar{L}_{k},\bar{A}_{k-1},
\bar{Z}_{k-1}, L_0) \) and
\(g_{Z_k} ( z \mid \bar{D}_k, \bar{Y}_k, \bar{C}_k,
\bar{L}_{k},\bar{A}_{k-1}, \bar{Z}_{k-1}, L_0) = P(Z_k = z \mid
\bar{D}_k, \bar{Y}_k, \bar{C}_k, \bar{L}_{k},\bar{A}_{k-1},
\bar{Z}_{k-1}, L_0) \). Our hypothetical interventions correspond to
modifications of these distributions to emulate the hypothetical
scenarios of balanced concomitant medication across treatment arms,
and full adherence to randomized treatment. Throughout, we shall
use the notation \(g_{A_k}^*, g_{Z_k}^*\) for these modified
hypothetical distributions.

Within this framework, the intention-to-treat analysis involves
interventions only on baseline randomized treatment \(A_0\) and
thus only a modification of the distribution \(g_{A_0}\) into a
degenerate \(g_{A_0}^*\) with all mass in 0 (to reflect assignment to
placebo group) or 1 (to reflect assignment to active treatment group);
otherwise trial participants can freely stop their treatment. 
Enforcing adherence involves modification of all distributions
\(g_{A_k}\) into degenerate \(g_{A_k}^*\) with all mass in 1 or, to
reflect that the participant starts and stays on active treatment
throughout the trial; see also Table \ref{table:overview:contrasted:interventions}.
%
For the placebo group we note that the active
treatment is not available, so that the degenerate \(g_{A_k}^*\) with
all mass in 0 (to reflect never initiating study treatment) is
straightforward.  

To balance exposure to concomitant treatment in the contrasted groups,
hypothetical interventions \(g_{Z_k}^*\) are defined for each \(k\)th
concomitant treatment decision \(Z_k\). To mimic different scenarios
where these treatment decisions are made without considering
randomized treatment information and time-dependent covariate
history, the interventional distributions \(g_{Z_k}^*\) are defined so
that they do not depend on these factors.
%
%
%
Table \ref{table:overview:drop-in:interventions} gives an overview of
different such interventions that we will consider. The static
intervention
\( g^{*}_{Z_k} ( z \mid \mathrm{Pa}(Z_k)) = \1\lbrace z = 0 \rbrace \)
prevents concomitant treatment usage entirely among all subjects.
While this likely corresponds to the ideal scenario to emulate, it may
suffer severely from lack of support in real data settings. A
different static intervention can be defined to enforce use of
concomitant treatment among all subjects,
\( g^{*}_{Z_k} ( z \mid \mathrm{Pa}(Z_k)) = \1\lbrace z = 1 \rbrace
\), but again plausible suffering from similar support issues.
To help improve this, a dynamic intervention can instead be defined
enforcing subjects exposed to concomitant medication at baseline
(\(Z_0 = 1\)) to continue taking the concomitant medication, and
preventing those not exposed at baseline (\(Z_{0}=0\)) from ever
initiating. Finally, to more closely imitate what was actually taken
in the trial, we also consider is a stochastic baseline covariate
dependent intervention that substitutes the distribution of \(Z_k\)
with one marginalized over the post-baseline covariates. This approach
indeed emulates a scenario where treatment decisions are not based on
post-baseline covariate information, and are instead made as if for an
average participant in the trial with similar baseline covariates and
similar history of concomitant treatment usage
.
As highlighted in Section \ref{sec1}, we again emphasize that, in
order to isolate the direct impact of randomized treatment from
its indirect impact through concomitant treatment (Figure
\ref{fig:intro:simple:dag}), this stochastic intervention is defined
unconditional on post-baseline covariates affecting the decision to
change usage of concomitant treatment.




\begin{table}
\begin{center}
\begin{tabular}{ r c c  l}
 \textit{Active arm adherence-enforcing intervention:} &  \(g^{*1}_{A_k} ( a \mid \mathrm{Pa}(A_k)) = \1\lbrace a
                                                                                                          = 1\rbrace \). \\
  \textit{Control arm no initiation enforcing intervention:} &  \(g^{*0}_{A_k} ( a \mid \mathrm{Pa}(A_k)) = \1\lbrace a
                                                                                                          = 0\rbrace\). 

\end{tabular}
\caption{Intention-to-treat and adherence-enforcing interventions on
  the randomized
  treatment.}\label{table:overview:contrasted:interventions}
\end{center}
\end{table}

\begin{table}
\begin{center}
\begin{tabular}{ r l l l}
  \textit{Static `\(z=0\)':}  & \( g^{*}_{Z_k} ( z \mid \mathrm{Pa}(Z_k)) = \1\lbrace z
                                = 0 \rbrace \) &   \textit{Prevent exposure to concomitant treatment.} \\
  \textit{{Static  `\(z=1\)':}}  & \( g^{*}_{Z_k} ( z \mid \mathrm{Pa}(Z_k)) = \1\lbrace z
                                        = 1 \rbrace \) &   \textit{Enforce exposure to concomitant treatment.} \\
  \textit{Dynamic:}  &  \( 
                       g^{*}_{Z_k} ( z \mid \mathrm{Pa}(Z_k)) = \1\lbrace z
                       = Z_{0} \rbrace \)  & \textit{Enforce continued use for those exposed at} \\
                              && \textit{baseline, prevent initiation otherwise. } \\
  \textit{Stochastic:}  & \( g^{*}_{Z_k} ( z \mid \mathrm{Pa}(Z_k)) =  \hat{P}_n (Z_k = z \mid   D_k,Y_k,\bar{Z}_{k-1}, L_0
                          ) \) &   \textit{Take concomitant medication as an average} \\
                              && \textit{participant in the trial with similar baseline} \\
                              & & \textit{covariates.} \\
\end{tabular}
\caption{Different interventions to balance exposure to concomitant
  medication post-baseline.
}\label{table:overview:drop-in:interventions}
\end{center}
\end{table}

\subsection{Target estimands}
\label{sec:target:estimand} 

For each set of hypothetical interventions proposed in Table
\ref{table:overview:drop-in:interventions}, we define correponding
causal estimands representing the effect measure of interest under the
corresponding post-interventional distributions.
Along with the treatment interventions discussed in Section
\ref{sec:hypothetical:interventions}, all post-interventional
distributions that we consider will further involve prevention of loss
to follow-up, corresponding to substituting \(g_{C_k}\) by the
degenerate
\(g_{C_k}^* (c \mid \mathrm{Pa}(C_k) ) = \1\lbrace c=0\rbrace \)
putting all mass in staying uncensored.
The interventions are carried out directly on the observed data
distribution factorized according to the observed variable ordering,
by substituting the relevant factors concerning randomized
treatment \(A_k\), concomitant treatment \(Z_k\) and censoring status
\(C_k\) across time by intervention\hyp{}specific choices.  This
corresponds to the substitution \(g\mapsto g^{*}\) by an
interventional choice \(g^{*}\) in the factorization \eqref{eq:P:QG},
and the resulting post-interventional distribution
\(P^{*} = P_{q,g^{*}}\) is what is generally referred to as the
g-computational formula \cite{robins1986new,gill2001causal}.
We then define each estimand as a contrast of expectations under the
post-interventional distribution \(P^{*1}=P_{q,g^{*1}}\), representing
the hypothetical active treatment arm (enforcing adherence to the
randomized treatment), and a post-interventional distribution
\(P^{*0}=P_{q,g^{*0}}\) (enforcing no initiation of randomized
treatment), representing the hypothetical control arm, with
\begin{align*}
  g^{*1} &= ( g_{A_k}^{*1}, g_{Z_k}^*, g_{C_k}^* \, : \, k=1,\ldots, K-1 ), \\
  g^{*0} &= ( g_{A_k}^{*0}, g_{Z_k}^*, g_{C_k}^* \, : \, k=1,\ldots, K-1 ) ,
\end{align*}
for each variant of the balancing interventions \(g_{Z_k}^*\) from
Table \ref{table:overview:drop-in:interventions}. We consider
specifically the risk differences
\begin{align}
  \psi = \EE_{P^{*1}} \big[ Y_{\KK}\big] - \EE_{P^{*0}} \big[
  Y_{\KK}\big],
  \label{eq:statistical:target:parameter}
\end{align}
comparing the hypothetical active treatment intervention to the
hypothetical control intervention evaluated at a fixed time-horizon
\(\KK\). 
In Section \ref{sec:identifiability} below, we discuss
assumptions under which the parameters defined by
\eqref{eq:statistical:target:parameter} represent the causal effect we
would had seen had we in fact carried out the hypothetical
interventions.


\subsection{Causal estimands}
\label{sec:causal:parameter}

For the purpose of discussing causal interpretability of our target
estimands defined by Equation \eqref{eq:statistical:target:parameter},
we introduce here counterfactual outcomes as follows: let
\(Y_{\KK}^{*1}\) denote the primary outcome status at time \(\KK\) we
would had observed for a subject had they been treated according to
\(g^{*1}\), and, likewise, let \(Y_{\KK}^{*0}\) denote the primary
outcome status we would had observed for the same subject had they
been treated according to \(g^{*0}\). For a similar definition of
counterfactuals, we refer to Gill and
Robins\cite{gill2001causal}. Under causal assumptions
, as discussed in Section \ref{sec:identifiability} below, we have that
\begin{align}
\psi = \EE [Y_{\KK}^{*1}] - \EE
  [Y_{\KK}^{*0}],
  \label{eq:causal:estimand}
\end{align}
i.e., the estimand defined by \eqref{eq:statistical:target:parameter}
represents the effect we would had seen, had we actually carried out
the hypothetical interventions. We can also write
\eqref{eq:causal:estimand} as
\begin{align*}\psi = \EE [T^{*1} \le t_{\KK},
  \Delta^{*1} = 1] - \EE [T^{*0} \le
  t_{\KK}, \Delta^{*0} = 1], 
\end{align*}
with \((T^{*1},\Delta^{*1}), (T^{*0},\Delta^{*0})\) denoting the pairs
of uncensored event time and event indicator we would had seen in the
active treatment arm and the control arm, respectively.

\subsection{Identifiability assumptions}
\label{sec:identifiability}


Two main conditions are necessary for our statistical estimands to
identify the corresponding causal estimands: no unmeasured confounding
(Assumption \ref{ass:nuc}) and positivity (Assumption
\ref{ass:pos}). Additional to these, it is further needed that there
is no interference among participants of the trial, in the sense that
changing the treatment for one should not affect the outcome for the
other. In the Supplementary Material, we repeat the proof 
of identification for completeness.


\begin{assumption}[No unmeasured confounding]
For the censoring mechanism,
it is required that
\begin{align}
  (Y_{\KK}^{*1},Y_{\KK}^{*0}) \independent
  C_k \mid \bar{L}_{k-1}, \bar{A}_{k-1}, \bar{Z}_{k-1},
  \bar{Y}_{k-1}=\bar{D}_{k-1}= \bar{C}_{k-1}=0,  \, \text{ for }\, k \le \KK-1, 
  \label{ass:nuc:cens}
  \intertext{and for the  randomized and concomitant treatment decisions that}
  (Y_{\KK}^{*1},Y_{\KK}^{*0}) \independent
  (A_k, Z_k) \mid \bar{L}_{k}, \bar{A}_{k-1}, \bar{Z}_{k-1},
  \bar{Y}_{k}=\bar{D}_{k}= \bar{C}_{k}=0,  \, \text{ for }\, k \le \KK-1.
  \label{ass:nuc:treat}
\end{align}
\label{ass:nuc}
\end{assumption}

\begin{assumption}[Positivity]
  Positivity is the requirement that
  \begin{align*}
   \sup_{o \in\mathcal{O}} \, \frac{
    \prod_{l=1}^{K} {g^*_{C_l} (c_l \mid \mathrm{Pa}(C_l))}{
    g^*_{Z_l} (z_l  \mid \mathrm{Pa}(Z_l)) g^*_{A_l} (a_l  \mid \mathrm{Pa}(A_l))}
    }{\prod_{l=1}^{K} {g_{C_l} (c_l \mid \mathrm{Pa}(C_l))}{
    g_{Z_l} (z_l  \mid \mathrm{Pa}(Z_l)) g_{A_l} (a_l  \mid \mathrm{Pa}(A_l))} }> \delta >0
  \end{align*}
  holds \(P_{0}\)-a.s., where \(\mathcal{O}\) is the support of \(O\). \label{ass:pos}
\end{assumption}

\begin{figure}[h]
  \vspace{-0.75cm}
  \centering \includegraphics[width=0.7\textwidth,angle=0]
  {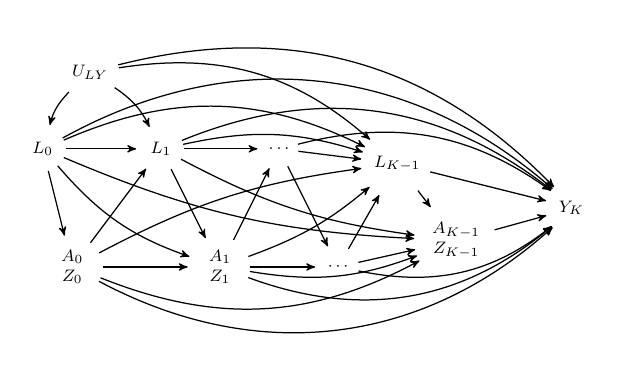} \vspace{-0.75cm}
  \caption{Simplified causal diagram to show the dependencies among
    treatment and covariates across time.  Here \(U_{LY}\) is a
    collection of unmeasured variables allowed to affect covariates
    and outcome across time, but not treatment decisions.  }
\label{fig:simple:dag} 
\end{figure}

Assumption \ref{ass:nuc} on no unmeasured confounders requires that
all factors influencing treatment decisions and censoring, which are
also predictive of outcome events, are measured.
Since the adherence-enforcing and balancing interventions are applied
to treatment decisions regarding randomized and concomitant medication
usage across all longitudinal time points, we need observation of all
relevant factors contributing to each of these decisions. See Figure
\ref{fig:simple:dag} illustrating the sequential nature of the problem
with a focus on the feedback between treatments and confounders across
time.  
%
For antidiabetic treatments, commonly measured confounders include
long-term blood glucose levels (e.g., HbA1c), the presence of
diabetes-related complications, and other relevant comorbidities
routinely considered in clinical decision-making. The completeness and
accuracy of trial data, particularly for these standard-of-care
variables, is critical to the plausibility of no unmeasured
confounding.

Assumption \ref{ass:pos} on positivity dictates that, at each
time-point \(k\), all trial participants who adhered so far to the
hypothetical treatment strategy should have a non-zero probability of
receiving any of the treatment options dictated by the interventions,
conditional on covariates. This is crucial both for the causal
interpretation of our target estimands and also for the validity and
performance of our statistical estimation procedure.
In the observed trial, randomization ensures that every participant
has a positive probability of being assigned to either the active
treatment or placebo, regardless of their characteristics, thereby
naturally satisfying the positivity assumption for intention-to-treat
interventions.
In the context of time-varying interventions, the situation becomes
more complex: over time, participants may stop using the
randomized treatment or start concomitant treatment for various
reasons, and, for some individuals, the probability of, e.g., not
initiating concomitant treatment throughout the follow-up period may
be very small.
%
%
Positivity violations occur when certain
subgroups rarely or never receive the treatment specified by the
hypothetical interventions. For example, individuals with poor
glycemic control (likely overrepresented in the control arm) may have
a very small probability of adhering to this treatment throughout
follow-up. Over time, the pool of participants adhering to both the
randomized treatment and no concomitant medication may shrink to
the point of being depleted. When this happens, the causal
interpretation is no longer valid, and the statistical estimation
procedure may suffer due to a lack of information for these
individuals.
%
%
%
%
The static, dynamic, and stochastic concomitant treatment
interventions outlined in Table
\ref{table:overview:drop-in:interventions} differ in how they assign
treatment and how stringent the corresponding positivity assumption
is. The static interventions apply the same treatment to all
individuals, regardless of their likelihood to initiate or maintain
concomitant medication, making them more susceptible to positivity
violations if, for example, no individuals stop taking concomitant
treatment after starting it. In contrast, the dynamic intervention
takes baseline treatment usage into account, enforcing treatment only
for those already taking it and withholding it for others, reducing
the risk of positivity violations. The stochastic intervention is
arguable more flexible, assigning treatment based on the likelihood of
individuals' treatment choices given their baseline characteristics,
further mitigating potential violations.

%

We further highlight how the impact of near-violations of the
positivity assumption varies depending on these different choices of
concomitant treatment interventions. These variations directly affect
the statistical estimation procedure through the so-called clever
weights (see Equation \eqref{eq:clever:weight} in Section
\ref{sec:ltmle}).
To generally satisfy positivity, deviations from clinical guidelines
(e.g., initiating glucose-lowering treatments based on HbA1c
thresholds) are required: such deviations are what allow
inference about hypothetical scenarios (e.g., what would happen to
individuals taking concomitant medication if they did not) but can
lead to extremely large clever weights in the estimation procedure for
individuals with very low probabilities of following the hypothetical
treatment.
These large weights can destabilize the estimation procedure and make it
highly sensitive to a small number of individuals.
%
%
The dynamic and stochastic interventions help address this issue by
limiting the magnitude of the weights assigned to individuals with
very low probabilities of following the hypothetical treatment. The
dynamic intervention adjusts the weights based on baseline treatment
usage, assigning greater weight to individuals who are not taking
concomitant medication (or already taking it) at baseline. The
stochastic intervention goes a step further, assigning weights based
on the overall likelihood of an individual’s treatment choices given
their baseline characteristics. Both help spread the influence across
covariate regions with larger support, eﬀectively downweighting the
contribution of individuals with very low probabilities of adhering to
the hypothetical treatment.

\section{Statistical estimation procedure}
\label{statistical:estimation}



To estimate the target estimands, we apply longitudinal targeted
minimum loss-based estimation estimation (TMLE)
\cite{van2012targeted,petersen2014targeted,ltmleRpackage}.  We briefly
outline the estimation procedure here and refer readers to the
Supplementary Material for a detailed description of the
implementation for our specific setting with competing risks and
stochastic interventions. Our focus is on providing the necessary
technical background for users, highlighting which aspects require
model specification or algorithmic choices. The estimation procedure
requires the following inputs, which we describe in more detail below:
\begin{enumerate}
\item Choice of estimation method for the sequential regression
  procedure across all longitudinal time-points \(k\); see Section
  \ref{sec:ice}.
\item Choice of estimation method for estimating distributions
  \(g_{Z_k}, g_{A_k}, g_{C_k}\) of all intervention variables
  \(Z_k, A_k, C_k\) across all longitudinal time-points \(k\). 
\end{enumerate}
\noindent The longitudinal TMLE takes 1. and 2. above as input and
applies sequentially a two-step procedure across each longitudinal
time-point \(k\) to provide a double robust and asymptotically
efficient estimator for each intervention-specific mean outcome of
\eqref{eq:statistical:target:parameter}, also when data-adaptive
machine learning tools are employed
\cite{van2011targeted,van2018targeted}. This sequential procedure is
specifically designed to correctly and optimally handle the complex
feedback between treatments and confounders across time as illustrated
in Figure \ref{fig:simple:dag}.

We give an overview of the longitudinal TMLE
procedure along with its theoretical properties in Section
\ref{sec:ltmle}, after briefly introducing the iterated conditional
expectations representation of the intervention-specific mean outcomes
in Section \ref{sec:ice}. In Section \ref{sec:super:learning} we
return to the choice of estimation methods for 1. and 2.

\subsection{Iterative sequence of conditional expectations}
\label{sec:ice}

The longitudinal TMLE procedure exploits the fact that each
intervention-specific mean outcome of
\eqref{eq:statistical:target:parameter} can be rewritten in terms of
an iterative sequence of conditional expectations
\cite{bang2005doubly}. These are defined for each longitudinal
time-point \(k\) as
\begin{align*}
  \bar{Q}_{k} (\bar{O}_{k-1})=
  \EE_{P_{q,g^*}}[ Y_{\KK}  \mid \bar{O}_{k-1} ]=
  \EE_{P}[ \underbrace{\EE_{P_{q,g^*}}[ \bar{Q}_{k+1} (\bar{O}_{k})  \mid {L}_{k}, {Y}_{k}, {D}_{k}, \bar{O}_{k-1} ]}_{
  = \bar{Q}^*_{k+1} ( {L}_{k}, {Y}_{k}, {D}_{k},\bar{O}_{k-1})} \mid \bar{O}_{k-1} ] ,
\end{align*}
where the right hand side shows how each \( \bar{Q}_{k}\) can be
obtained from \(\bar{Q}_{k+1}\) by first integrating out intervention
variables (\(Z_k, A_k, C_k\)) according to the specified
interventional distributions \(g_{Z_k}^*, g_{A_k}^*, g_{C_k}^*\)
(inner expectation of the right hand side), to obtain
\(\bar{Q}^*_{k+1} ( {L}_{k}, {Y}_{k}, {D}_{k},\bar{O}_{k-1})\), and
next integrating out non-interventional variables
(\({L}_{k}, {Y}_{k}, {D}_{k}\)) according to the observed-data
distribution. For time-point \(k=1\), we have that
\(\psi^{a'} = \EE_P [ \bar{Q}^{*}_1 (L_0) ]\) where the expectation is
taken over the (observed) marginal distribution of baseline covariates
\(L_0\).

We highlight that \(D_k\) is used in this process (along with \(Y_k\))
solely to remove patients from being at risk, and that survival
probabilities are not modelled directly. In the Supplementary
Material, we clarify that this yields the targeted interpretation of
intervention-specific mean outcomes as intervention-specific absolute
risk probabilities, appropriately accounting for the competing risk of
death.


\subsection{Longitudinal TMLE}
\label{sec:ltmle}

The longitudinal TMLE for estimation of a given intervention-specific
mean outcome proceeds in two steps at each longitudinal time-point
\(k\) as follows:
\begin{enumerate}
\item \textit{Regression step, taking
    \(\hat{\bar{Q}}^{\mathrm{tmle}}_{k+1,n}(\bar{O}_{k})\) as input. }
  This step is carried out to learn how the conditional outcome risk
  under the hypothetical interventions depends on the history up until time
  \(k-1\).  First the interventional variables are integrated out from
  \(\hat{\bar{Q}}^{\mathrm{tmle}}_{k+1,n}(\bar{O}_{k})\) to obtain
  \(\hat{\bar{Q}}^{*,\mathrm{tmle}}_{k+1,n}(L_k, Y_k, D_k,
  \bar{O}_{k-1})\).  Next
  \(\hat{\bar{Q}}^{*,\mathrm{tmle}}_{k+1,n}(L_k, Y_k, D_k,
  \bar{O}_{k-1})\) is regressed on \(O_{k-1}\).  The obtained
  estimator is denoted \(\hat{\bar{Q}}_{k,n}(\bar{O}_{k-1})\).
\item \textit{Targeting step, taking
    \(\hat{\bar{Q}}^{*,\mathrm{tmle}}_{k+1,n}(L_k, Y_k, D_k,
    \bar{O}_{k-1})\) and \(\hat{\bar{Q}}_{k,n}(\bar{O}_{k-1})\) as
    input.} In this step, the regression estimator
  \(\hat{\bar{Q}}_{k,n}(\bar{O}_{k-1})\) is updated using the product
  of estimated propensity scores and censoring probabilities across
  all time-points \(l=1,\ldots, k-1\), in a properly weighted
  intercept-only regression of
  \(\hat{\bar{Q}}^{*,\mathrm{tmle}}_{k+1,n}(L_k, Y_k, D_k,
  \bar{O}_{k-1})\) on offset
  \(\hat{\bar{Q}}_{k,n}(\bar{O}_{k-1})\). This specification of the
  model, and particularly the weights, is guided by the so-called
  \textit{efficient influence curve} of the given
  intervention-specific mean outcome. After this step, we say that the
  step-\(k\) regression estimator is \textit{targeted}, and is denoted
  by \(\hat{\bar{Q}}^{\mathrm{tmle}}_{k,n}(\bar{O}_{k-1})\).
\end{enumerate}
Carrying out this procedure across longitudinal time-points
\(k=K,\ldots, 1\) yields in the end an estimator for the
intervention-specific mean outcome,
\(\hat{\psi}^{\mathrm{tmle}}_n = \mathbb{P}_n
\hat{\bar{Q}}^{*,\mathrm{tmle}}_1 (L_0)\), which, by design of the
targeting step, solves the so-called \textit{efficient influence curve
  equation}. Particularly, inference can then be based on the
efficient influence curve when also further conditions
\cite{van2006targeted,van2017generally,van2011targeted}
related to the estimation of \(Q\) and \(g\) (the product of their
rates of convergence have to be faster than \(n^{-1/2}\)) hold true.

  Each intervention-specific mean outcome can be represented as
  \(\psi^{a'} =  \EE_{P^{*a'}} \big[ Y_{\KK}\big] = \Psi^{a'}(P)\), where
  \(\Psi^{a'} \, : \, \mathcal{M}\rightarrow \R\) has efficient
  influence curve given by
\begin{align}
  & \phi^{a'}( P) (O) = \sum_{l=1}^{\KK} \phi_l ( P) (O)
    + \bar{Q}^*_{1} ({L}_{0}) -  \EE_{P^{*a'}} [  Y_{\KK} ],
    \label{eq:efficient:ic}
\end{align}
with the \(l\)-specific components defined for \(l=1,\ldots, \KK\) by
\begin{align*}
  & \phi^{a'}_l ( P) (O) =
    H_l (g) (\bar{O}_{l-1})  \big(  \bar{Q}^*_{l+1}
    (\bar{O}_{l-2}, {L}_{l-1}, {Y}_{l-1}, {D}_{l-1})
    -  \bar{Q}_{l} (\bar{O}_{l-1}) \big) , 
\end{align*}
and weights \(H_k (g) (\bar{O}_{k-1})\)  defined as 
\begin{align}
  H_k (g) (\bar{O}_{k-1}) & = \1 \lbrace Y_{k-1}=D_{k-1}=0\rbrace H^{\mathrm{cens}}_k (g) (O)  H^{\mathrm{treat}}_k (g) (\bar{O}_{k-1}), \label{eq:clever:weight}
                            \intertext{with}
                            H^{\mathrm{cens}}_k (g) (\bar{O}_{k-1}) 
  &=
    \prod_{l=1}^{k-1} \frac{g^*_{C_l}( C_l \mid \mathrm{Pa}(C_l)) }{g_{C_l} (C_l  \mid \mathrm{Pa}(C_l))}
    =
    \prod_{l=1}^{k-1} \frac{\1 \lbrace C_l =0 \rbrace }{g_{C_l} (0 \mid \mathrm{Pa}(C_l))} ,  \notag
    \intertext{and}
    H^{\mathrm{treat}}_k (g) (\bar{O}_{k-1})
  &=
    \prod_{l=1}^{k-1} \frac{g^*_{Z_l}( Z_l \mid \mathrm{Pa}(Z_l) )g^*_{A_l}( A_l \mid \mathrm{Pa}(A_l) )}{
    g_{Z_l} (Z_l  \mid \mathrm{Pa}(Z_l)) g_{A_l} (A_l  \mid \mathrm{Pa}(A_l))}. \notag
\end{align}
We note that the structure of the efficient influence
curve presented in \eqref{eq:efficient:ic} corresponds to that
presented for general longitudinal data structures in Appendix A.7 of
van der Laan and Rose 2011\cite{van2011targeted}; the only difference
is the clever weights that we defined in \eqref{eq:clever:weight}
which for our setting includes indicator functions tracking the
survival and event status of subjects.
The asymptotic distribution of the targeted estimator
\( \hat{\psi}^{a',\mathrm{tmle}}_n \) is achieved from its
asymptotically linear representation
\begin{align}
  \sqrt{n} \big(  \hat{\psi}^{a',\mathrm{tmle}}_n - \psi^{a'}_0 \big) = \sqrt{n} \,
  \mathbb{P}_n \phi^{a'}(P_0) + o_P(1),
  \label{eq:asymptotic:distr}
\end{align}
which implies directly that \(\hat{\psi}^{a',\mathrm{tmle}}_n\)
follows an asymptotic normal distribution centered around the
corresponding true value \(\psi^{a'}_0 =\Psi^{a'}(P_0)\) and with
variance given by the variance of the efficient influence curve.
Specifically, the variance of the estimator can be estimated by
\(\hat{\sigma}_n^2 /n\) where
\(\hat{\sigma}_n^2 = \mathbb{P}_n (\phi^{a'} (
\hat{P}^{a',\mathrm{tmle}}_n))^2\). Inference provided for estimators
presented in the simulation study and the data analysis are all based
on the asymptotic distribution displayed in
\eqref{eq:asymptotic:distr}.

\subsection{Super learning}
\label{sec:super:learning}

The TMLE procedure requires estimation of the regression steps, to
learn how each
\( \bar{Q}^*_{k+1} ( {L}_{k}, {Y}_{k}, {D}_{k},\bar{O}_{k-2})\) and
\(\bar{Q}_{k} (\bar{O}_{k})\) \(\bar{Q}_{k} (\bar{O}_{k-1})\) are
related, and further estimation of distributions
\(g_{Z_k}, g_{A_k}, g_{C_k}\) of all intervention variables
\(Z_k, A_k, C_k\) across all longitudinal time-points \(k\).
We emphasize that it is implausible that we are able to correctly
specify parametric regression models for all these different
factors. This highlights the benefits of using super learning
\cite{van2007super,polley2011super} to flexibly learn how follow-up
treatment decisions and outcome risk depend on previous health
measurements and treatment usage. Super learning allows us to combine
different estimation methods (including machine learning algorithms
and/or different regression models), that we need to specify in a
library: the super learner then picks out, or optimally weights,
predictions obtained with the different estimation methods using cross
validation. Importantly, the inferential properties of the resulting
TMLE estimators \eqref{eq:asymptotic:distr} are still valid under the
conditions that super learner library specified either contains a true
parametric model or a machine learning algorithm which is consistent
at a fast enough rate.
For the regression steps, any method that can handle outcomes in the
interval \([0,1]\), such as a quasi-logistic regression or random
forests, can be included in the super learner library.  For the
conditional distributions of binary variables, any binary regression
method applies. 

\color{black}

\section{Simulation study}
\label{sec:simulation:study}

We consider a simulation study to illustrate estimation of the
estimands defined in Section \ref{sec:causal:questions} under
different scenarios with varying levels of concomitant medication
usage.  We design the simulation study to loosely imitate the data
analysis setting (Section \ref{sec:leader:analysis}) with unbalanced
initiation of glucose-lowing medications in cardiovascular outcomes
trials.  To keep things simple, and to focus exclusively on the
choices of hypothetical interventions for defining our estimands of
interest, we simulate from main effects parametric regression
models. The simulated data includes a time-dependent covariate \(L_k\)
which overall increases the risk of occurrence of primary outcome
\(Y_k\).  We do not consider competing risks or right-censoring.

At baseline, we draw variables 
$$ \quad L_0 \sim N(0,1),\quad P(A_0=1) = 0.5,
\quad P(Z_0=1\mid L_0) = \mathrm{expit} (L_0 + c_{Z0}),$$ where
\(c_{Z0}\in \R\) is a parameter that shifts the overall rate of
concomitant medication use at baseline.  Patients with high values of
$L_0$ are more likely to have $Z_0 = 1$ in this model.  In the context
of the trials, we can interpret \(L_0\) and $L_k$ as the (scaled)
logarithm of positive and bounded from below lab measurements such as
HbA1c.

{During follow-up, both randomized treatment and concomitant
  treatment usage decrease the risk of randomized outcome
  occurrence and levels of the time-dependent covariate.} We use the
running average of treatment, defined as
\(\bar{\bar{A}}_k = \frac{1}{k+1} \sum_{l=0}^k A_l\), to summarize
treatment history, and define \(\bar{\bar{Z}}_k\) and
\(\bar{\bar{L}}_k\) similarly. This can easily be extended to a
discounted average.  However, to focus on the interventions to handle
unbalanced concomitant treatment usage across randomized
treatment arms, we assume patients are fully adherent to the
randomized treatment (i.e. $A_k = A_0$), so the running average
satisfies $\bar{\bar{A}}_k = A_0$. The time-dependent covariates are
simulated as
\begin{equation}
L_k \sim N(m_k, 0.5^2), \quad m_k = L_{k-1} - 0.3 (\bar{\bar{A}}_{k-1} +p_{Z}
\bar{\bar{Z}}_{k-1}).
\end{equation} 
Here, \(p_Z\ge 0\) measures the efficacy of the concomitant treatment
in reducing values of \(L_k\) relative to the randomized
treatment.

For the outcome, $Y_{k-1} = 1$ implies $Y_k = 1$
deterministically. Conditional on survival, we simulate the outcome \(Y_k\)
during follow-up according to the model
\begin{equation}
  P(Y_k= 1 \mid Y_{k-1} = 0,  \bar{L}_{k-1},
  \bar{A}_{k-1}, \bar{Z}_{k-1}) =
  \mathrm{expit} \big( 0.3(\bar{\bar{L}}_{k-1} - 
  \bar{\bar{A}}_{k-1} - p_{ZY} \bar{\bar{Z}}_{k-1} ) -3.75 \big).
\end{equation} 
The overall baseline rate is governed by the intercept (equal to
$-3.75$ here), and $p_{ZY}$ measures the relative efficacy of the
concomitant treatment in reducing the risk of the outcome. Note that
we have assumed the relative efficacy is the same for both the
covariates and outcome.

To model the imbalance in concomitant initiation in the two arms,
concomitant treatment use for visit $k$ is simulated from
\begin{equation}
  P(Z_k = 1 \mid 
  Y_{k-1} = 0,   \bar{L}_{k},
  \bar{A}_{k-1}, \bar{Z}_{k-1} ) = \text{expit}( L_{k} + 8Z_{k-1} +c_Z),
\end{equation}
where $b_{ZZ} =8$ controls the probability of remaining on the
concomitant treatment and $c_Z$ controls the overall rate of
concomitant initiation. Imbalance in concomitant usage occurs as $L_k$
will decrease less noticeably in the placebo arm, leading to a higher
probability of $Z_k =1$ in the placebo arm.  Note that $Z_{k-1}$ is
multiplied by a large constant (equal to 8 here) so patients who
initiate concomitant will be very unlikely to discontinue.  We
highlight that the initiation of concomitant $Z_k$ is only affected by
the main treatment $A_k$ indirectly through the lowering of the
time-varying confounder $L_k$. As we have argued, this is a reasonable
assumption under a blinded trial.

The true value of each estimand is approximated with Monte Carlo by
simulating from the respective counterfactual scenarios, i.e.,
replacing observed data mechanisms for $\{A_k, Z_k\}$ by their
interventional distributions. We focus on estimation of the three
classes of estimands as introduced in Table
\ref{table:overview:drop-in:interventions}, and compare to an `ignore
concomitant' estimand where concomitant treatment usage is not intervened
on.

We consider three different simulation scenarios, corresponding to
different settings of the parameters $\{c_{Z0},c_Z, p_Z\}$. For all
scenarios, we simulate $K_0 = 5$ visits for $n = 9340$ patients, and
repeat over 2000 different seeds to evaluate coverage. For targeted
learning estimation, we elicit logistic regression models for the
initial outcome and propensity score models adjusting for the relevant
variables.

\subsection{Results}

Table \ref{tab:scenarios:params} illustrates three different
simulation scenarios. Scenario 1 is the reference scenario, with low
concomitant treatment initiation rate and concomitant treatment that
is of equal efficacy relative to the randomized treatment.
Scenario 2 considers the case where the concomitant treatment
initiation rate is higher. Finally, Scenario 3 is the same as Scenario
1 except the efficacy of the concomitant treatment is much weaker than
the randomized treatment. For reference, the baseline rate of the
outcome at the final visit under the static intervention with
$(a = 0, z = 0)$ is equal to 11.4\%.

\begin{table}[!h]
\centering
\vspace{2mm}
\begin{tabular}{c|c|c|c}
  Scenario & Efficacy & Drop-in rate &  Comments    \\
  \hline
  1 & $p_Z = 1$ & $c_{Z0} = -1.5,\, c_Z = -2.5$ &   Low  rate of concomitant treatment usage \\
  2& $p_Z = 1$ & $c_{Z0} = -1,\, c_Z = 0$   &  High rate of concomitant treatment usage \\
  3& $p_Z = 0.1$ &  $c_{Z0} = -1.5,\, c_Z = -2.5$& Low  efficacy  of concomitant treatment \\
\end{tabular}\caption{Scenarios with parameter settings.}
\label{tab:scenarios:params} 
\end{table}

\begin{figure}[!h] 
  \centering \includegraphics[width=0.9\textwidth,angle=0]
{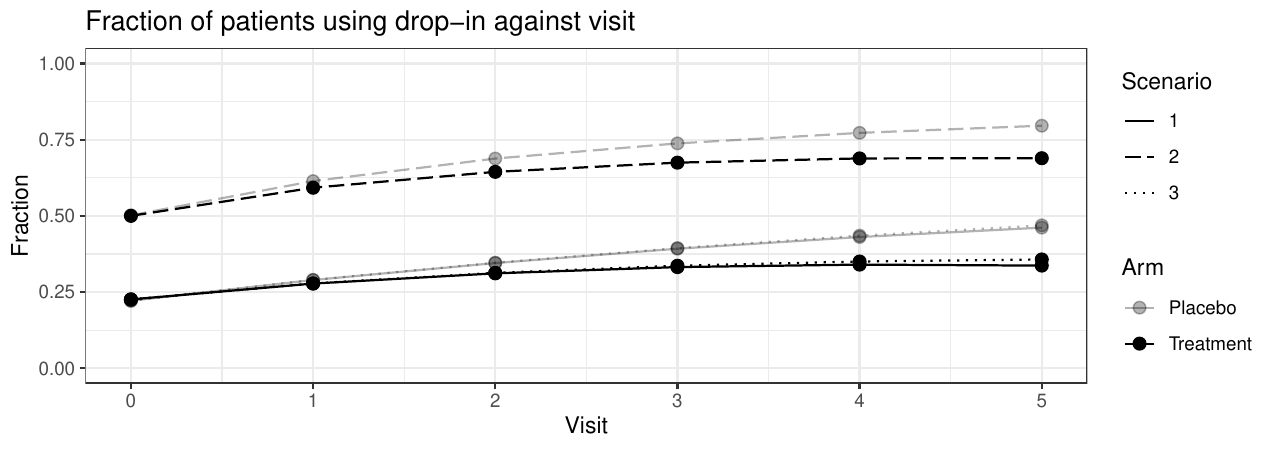}
\vspace{-2mm}
\caption{Fraction of drop-in in each arm for the three
  scenarios.}
\label{fig:simulation:traj}
\end{figure}

Figure \ref{fig:simulation:traj} illustrates the marginal rate of
concomitant use among the population for the three scenarios. In the
default Scenario 1, we see that the usage of concomitant treatment
increases noticeably more in the placebo arm versus the treatment arm,
driven by the higher values of $L_k$ in the placebo arm. 
In Scenario 2, we have a higher initial rate of concomitant treatment
usage as well as a higher probability of initiating, with a similar
separation of the placebo and treatment arm. Finally, we note that in
Scenario 3, the trajectories of concomitant treatment usage is mostly
unchanged by reducing $p_Z$ compared to Scenario 1, thus pointing to
the randomized treatment as the main driver in concomitant usage
trajectories.

\begin{figure}[!h] 
  \centering \includegraphics[width=\textwidth,angle=0]
{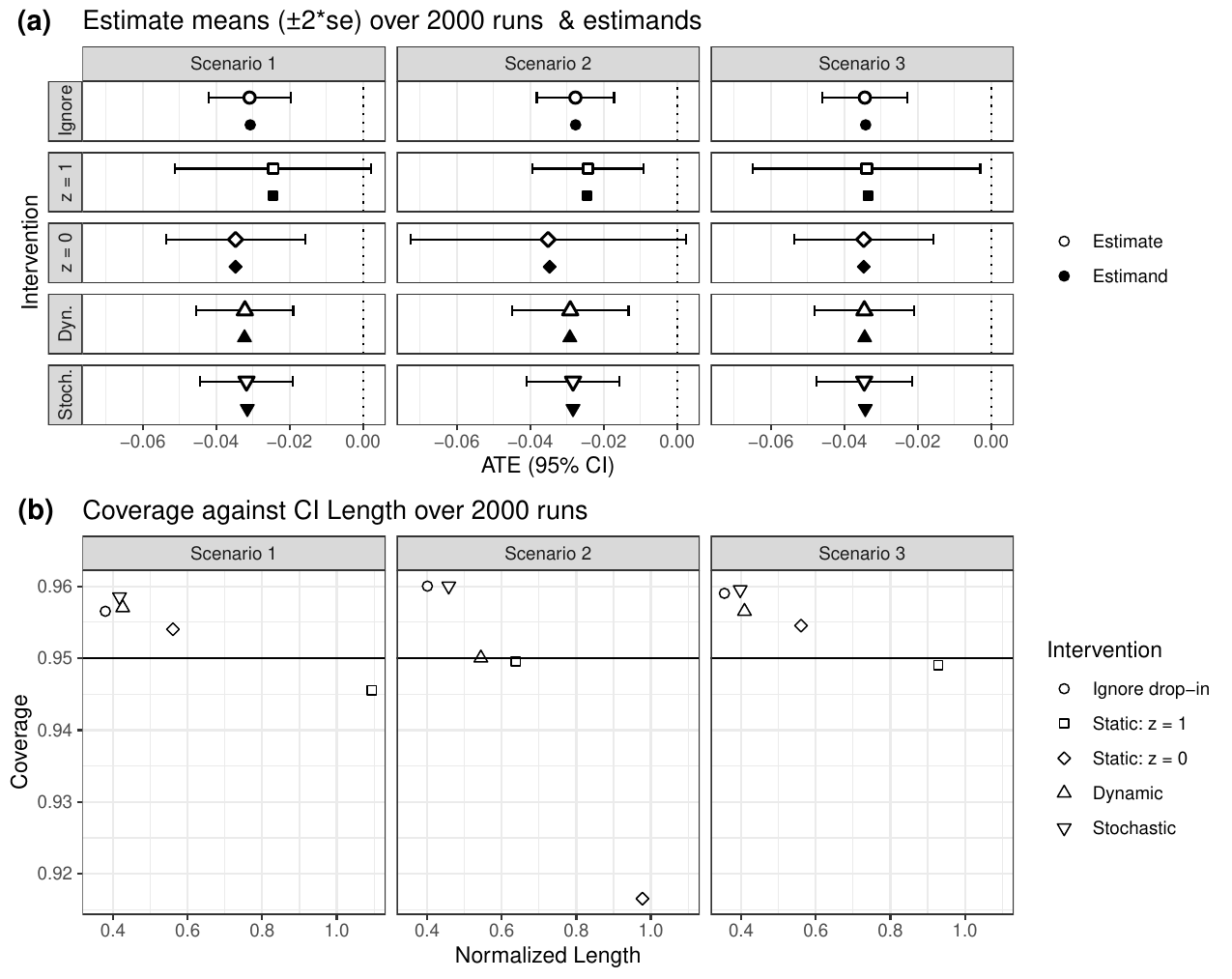}
\caption{Simulation results for 2000 runs with (a) estimands and
  estimates (mean and $\pm1.96$ standard deviations); (b) coverage
  against normalized CI lengths; Ignore [concomitant treatment usage]
  = no interventions on concomitant treatment usage; Static, Dynamic,
  Stochastic and defined in Table
  \ref{table:overview:drop-in:interventions}. }
\label{fig:simulation:results}
\end{figure}

Figure \ref{fig:simulation:results}(a) illustrates the true value of
the estimands under the different interventions for the three
scenarios. We see that in Scenario 1, the effect measure corresponding
to the static intervention enforcing no concomitant treatment usage
has the largest magnitude, which is unsurprising as the baseline risk
is also the highest. Analogously, the estimand corresponding to a the
opposite static intervention enforcing concomitant treatment usage
results in the effect measure with the smallest magnitude. The
estimand for which concomitant treatment usage is ignored as well as
the estimands corresponding to the dynamic and stochastic
interventions lie between the static interventions, as roughly
35--50\% of patients will be using the concomitant treatment by the
end of trial. The true values of the estimands are the same in
Scenarios 2 and 1, as only the model for $Z_k$ is changed. Finally, in
Scenario 3, we note that the relative relationships between the
interventions are the same, but the spread between the static
interventions is much smaller. This is consistent with the lowering of
efficacy of the concomitant treatment. In summary, the differences in
estimand values for the different interventions depend both on the
level of imbalance in concomitant usage in the two arms and the
relative efficacy of the concomitant treatment.  Figure
\ref{fig:simulation:results}(a) also illustrates the mean (with
$\pm1.96$ standard deviation bars) of the LTMLE estimates over 2000
runs, which we can see are unbiased. We see clearly that the widths of
the 95\% confidence intervals for the static estimands vary noticeably
between Scenario 1 and 2 due to the differing concomitant rate, while
the widths for the dynamic and stochastic interventions stay mostly
unchanged.

Figure \ref{fig:simulation:results}(b) illustrates the coverage and
normalized lengths of the 95\% confidence intervals over 2000
runs. The length of the 95\% confidence intervals are divided by the
absolute value of the respective true estimand values for
comparability. Here we will primarily focus on contrasting Scenarios 1
and 2, which corresponds to low and high rate of concomitant treatment
usage. We see that in Scenario 1, the estimand ignoring concomitant
and the dynamic and stochastic estimands are close in coverage and
normalized length, with values noticeably smaller than the static
interventions. The static estimand for $z = 0$ attains slightly better
coverage but with much smaller relative lengths compared to $z=1$,
indicating that there is more support in Scenario 1 for the $z=0$
estimand. In Scenario 2 however, the above relationship is
reversed. The estimates for $z = 0$ have larger relative length and
lower coverage than $z=1$, which is unsurprising as the rate of
concomitant is much higher in Scenario 2, leading to less data support
for the $z= 0$ estimand. We highlight here that positivity and support
depend both on the concomitant rate and the evolution of the
time-varying confounder.  In all three scenarios however, the
stochastic intervention provides the most robust confidence intervals,
closely followed by the dynamic intervention, indicating that these
have the best support over a wide range of settings.

\section{Data analysis}
\label{sec:leader:analysis}

\begin{figure}[!h] 
  \centering \includegraphics[width=0.9\textwidth,angle=0]
{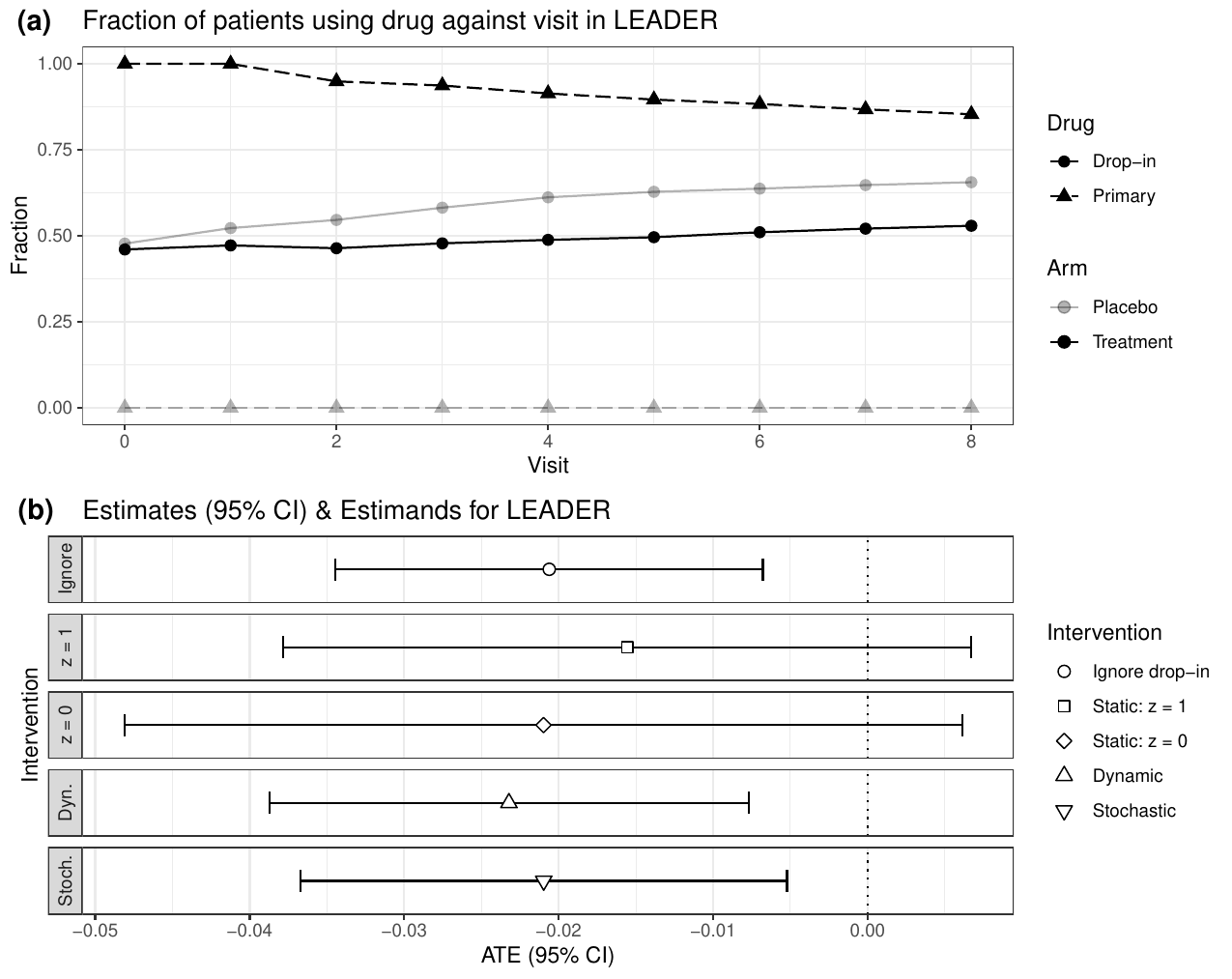}
\vspace{-2mm}
\caption{(a) Fraction of those at risk using main/drop-in treatment in each arm for LEADER and (b) LTMLE estimates with 95\% confidence intervals.}
\label{fig:leader:results}
\end{figure}

The LEADER trial was a long-term, multi-centre, international,
randomized cardiovascular outcomes trial including 9,340 patients with
type-II diabetes and high cardiovascular risk who were randomized to
either treatment with GLP1 (liraglutide) or placebo. Additional care
of subjects was otherwise decided by the subject's physician, and,
particularly, additional glucose-lowering medications such as insulin
were allowed to be added the treatment regimen. Subjects were followed
over time until death, or loss to follow-up, and attended follow-up
visits after randomization every six months up to a maximum of 10
follow-up visits, with an additional initial follow-up visit at the
3-month mark. See Marso et al.\cite{marso2016liraglutide} for further
details. For our analysis, we focus on insulin as our main example of
a concomitant treatment: of the 9,340 trial participants, 4,169 were
already taking insulin at baseline. Of the remaining 5,171
participants, 30\% initiated insulin at some point during follow-up in
the GLP1 arm whereas 47\% initiated insulin in the placebo arm. This
divergence in concomitant usage is illustrated in Figure
\ref{fig:leader:results}a, which matches quite closely to Scenario 1
in the simulations. %
During follow-up, 1,302 randomized events (time to first event of
cardiovascular death, non-fatal myocardial infarctions, or non-fatal
stroke) were observed, 270 deaths due to non-cardiovascular causes
occurred and 7,768 subjects were either event-free at end of follow-up
or lost to follow-up.

We included three time-varying covariates in our analysis: glycated
haemoglobin (HbA1c) which was measured at all visits, estimated
glomerular filtration rate (eGFR) and body mass index (BMI) which was
measured {at a subset of all visits} for each participant.  We
considered the trial visits as our time-grid at which changes to
observed processes can happen; if no update of a given covariate was
measured at a given visit, we carried forward the last known value. We
included sex, age, diabetes duration, history of cardiovascular
disease, BMI, HbA1c, eGFR and insulin history measured at the time of
randomization as baseline covariates.

To apply our methodology to the LEADER dataset, careful pre-processing
was necessary to avoid bias. In particular, concomitant treatment and
occurrence of an event/censoring/death are inherently continuous
processes, as the specific dates of initiation/discontinuation or an
event are reported. However, our method assumes discrete-time
processes, so discretization according to the trial visit time-grid is
necessary. At visit $k$, we define $Z_k = 1$ if the patient initiates
or continues to use insulin at \emph{any} time between visits $k-1$
and $k$.  Similarly, we set $Y_k$, $C_k$ or $D_k$ to 1 if a respective
event/censoring/death was observed between visits $k-1$ and $k$.
Finally, to ensure correct time-ordering among variables across time,
we define the time-varying covariate $L_{k}$ according to the
covariate measurement at the \emph{previous} trial visit $k-1$. This
is key, as failure to lag the covariate will result in the reverse
causal relationship of $Z_k \to L_k$. This is because $L_k$ occurs at
(or just around) visit $k$, whereas $Z_k$ is a summary of the window
from visit $k-1$ to $k$. In this specific context, this can be
particularly misleading as concomitant treatment
initiation/discontinuation often occurs directly after a covariate
measurement is taken.

\subsection{Results}

As in the simulations, we show results for the four classes of
interventions on the concomitant treatment. In addition, we apply only
the static interventions of $A_k = 1$ and $A_k =0$ to the
randomized treatment in the respective arms, yielding the `ignore
concomitant treatment usage' intervention. Our estimands of interest
are the absolute risk contrasts under the various interventions at the
eight visit, which corresponds to 42 months after randomization. For
initial estimation, we utilize a SuperLearner consisting of the
\texttt{glm}, \texttt{bayesglm} and \texttt{gam} libraries.

The estimates with 95\% confidence intervals are shown in Figure
\ref{fig:leader:results}b. The point estimate for the static
intervention with $z=1$ has the smallest magnitude, whilst the point
estimate for $z=0$ matches with the estimand ignoring concomitant
treatment usage. The point estimate for the dynamic intervention is
slightly larger in magnitude, and the stochastic intervention also
matches the estimand ignoring concomitant. The differences between the
four estimands are small because insulin treatment does not
substantially alter the risk of the randomized event. In terms of
the confidence intervals, a key observation is that the intervals for
both static interventions are wide, even crossing the boundary for
significance, which indicates a lack of data support. On the other
hand, both the dynamic and stochastic intervention estimates remain
significant, with the interval widths being only slightly wider than
the estimand ignoring concomitant. This highlights the need to be
careful when utilizing static interventions, and also emphasizes the
utility of the dynamic and stochastic interventions.

\section{Discussion}
\label{sec:discussion}


Information collected at trial visits in randomized controlled trials
offers opportunities for the evaluation of relevant treatment regimes
that change over time. In this paper we have proposed a class of
causal estimands to handle unbalanced drop-in of concomitantly taken
outcome-protective medication in randomized controlled trials. Each
causal estimand is indexed by a particular hypothetical time-varying
intervention which isolates the specific impact of the randomized
treatment as it would be observed had the use of the concomitant
medication, in one way or another, been balanced across the
randomization arms. The causal estimands are then defined as an
absolute risk evaluated at a fixed time-point under the respective
post-interventional distribution. For estimation, we have implemented
a longitudinal targeted minimum loss-based estimation procedure for
our particular estimands in the general setting considered
characterized by time-dependent confounding, terminal outcome events,
and competing risks.
In the same framework, we note that we could likewise define and
estimate effects representing the amount of treatment modification
there exists between the randomized treatment and concomitant
treatments.
We also emphasized, that while we focused on a single concomitant
medication, the methods could straightforwardly be extended to handle
multiple types of concomitant medications. Similarly, the censoring
process could be split into two types of censoring, tracking loss to
follow-up on the one hand, and administrative censoring on the other.
Our methods overcome key limitations of existing approaches by
enabling valid statistical inference for absolute risk under flexible,
data-adaptive adjustment for time-dependent confounding, without
relying on restrictive model assumptions or hazard ratios with
problematic causal interpretations.

Causal interpretability of the target estimands in the end relies on
the causal assumptions of positivity and no unmeasured
confounding. The proposed stochastic hypothetical interventions were
defined particularly to improve upon the plausibility of positivity:
in a randomized controlled trial the randomization groups are balanced
at baseline, so that unbalanced use of concomitant treatment occurs
not initially but may grow as a problem over the course of time.  The
stochastic interventions allow us to define a causal effect of the
randomized treatment under more realistic balanced use of
concomitant medication than dictated by a static intervention imposing
no use of concomitant treatment. {In our simulations, we found, as
  expected, that the static balanced interventions lacked
  positivity. } For the assumption of no unmeasured confounding we
need to measure all covariates that are both predictive of outcome and
of treatment changes as well as loss to follow-up during
follow-up. Even though the randomized trials are only secured by the
randomization at baseline, we still point out other advantages: the
quality of information that is collected at the trial visits about the
health status of participants and the absence of switches from placebo
to active treatment arm.

\section{Data availability statement} 
\label{sec:data:availability}

Research data are not shared.

\section*{Acknowledgements}

The project was carried out as part of the \textit{Joint Initiative of
  Causal Inference} (JICI) collaboration between the University of
California, Berkeley, the University of Copenhagen and Novo Nordisk
A/S. Funding was provided by a philanthropic gift from Novo Nordisk
A/S to the Section of Biostatistics at the University of Copenhagen
and to the Center for Targeted Machine Learning and Causal Inference
at the University of California, Berkeley.

\appendix


\section{Overview of notation}

\begin{table}[!h]
\begin{center}
\begin{tabular}{ r l}
  \(L_0\) & Baseline covariates. \\
  \(A_k\) & Randomized treatment at visit number \(k\)  \\
  \(Z_k\) & Concomitant treatment at visit number \(k\)  \\
  \(Y_k\) & Outcome status at visit number \(k\)  \\
  \(D_k\) & Competing death status at visit number \(k\)  \\
  \(C_k\) & Censoring status at visit number \(k\)  \\
\end{tabular}
\caption{Overview of notation for observed data
  variables.}\label{table:app:notation:variables}
\end{center}
\end{table}

\begin{table}[!h]
\begin{center}
\begin{tabular}{ r l}
  \(g_{A_k} ( a \mid \mathrm{Pa}(A_k)) = P(A_k = a \mid
\bar{D}_k, \bar{Y}_k, \bar{C}_k, \bar{L}_{k},\bar{A}_{k-1},
\bar{Z}_{k-1}, L_0)\)  &  conditional distribution of
randomized treatment status \(A_k\)  \\
    \(g_{Z_k} ( z \mid \mathrm{Pa}(Z_k)) = P(Z_k = z \mid
\bar{D}_k, \bar{Y}_k, \bar{C}_k, \bar{L}_{k},\bar{A}_{k-1},
\bar{Z}_{k-1}, L_0) \) &  conditional distribution of
concomitant treatment status \(Z_k\)   \\
  \(g_{C_k}( c \mid \mathrm{Pa}(C_k)) = P(C_k= c \mid
\bar{D}_k, \bar{Y}_k, \bar{C}_k, \bar{L}_{k},\bar{A}_{k-1},
\bar{Z}_{k-1}, L_0)\)  & conditional distribution of
randomized treatment status \(A_k\)
\end{tabular} 
\caption{Overview of notation for components of the interventional
  part of observed data
  distribution.}\label{table:app:notation:interventional}
\end{center}
\end{table}

\section{Statistical estimation procedure}

\subsection{Representation of target parameters by iterated conditional expectations}
\label{sec:representation:iterated:expectations}

Following Bang and Robins\cite{bang2005doubly}, the
intervention-specific mean outcome \(\psi\) can be identified through
a sequence of conditional expectations. We here {present} notation for
this representation, that specifically addresses competing risks and
stochastic interventions.  An overview of the notation is provided in
Table \ref{tab:overview:Q:notation}.  For the following, we fix a
specific set of interventions
\( g^{*} = ( g_{A_l}^{*}, g_{Z_l}^*, g_{C_l}^* \, : \, l=0,\ldots, K-1
)\).

First, for the time-point \(t_{\KK}\) at which the risk is evaluated,
we define \( \bar{Q}^*_{\KK+1} := Y_{\KK} \) and then as follows
\begin{align*}
  &  \bar{Q}_{\KK} (\bar{O}_{\KK-1})
    = \1 \lbrace Y_{\KK-1} = D_{\KK-1}=0\rbrace   
    \underbrace{\EE_P  [ \bar{Q}^*_{\KK+1} \mid \bar{A}_{\KK-1},\bar{Z}_{\KK-1}, \bar{L}_{\KK-1},
    \bar{Y}_{\KK-1}, \bar{D}_{\KK-1},  \bar{C}_{\KK-1}]}_{=: \tilde{\bar{Q}}_{\KK} (\bar{O}_{\KK-1})}
    \notag  
    + \1 \lbrace Y_{\KK-1}= 1\rbrace ,\notag
\end{align*}
where \(\tilde{\bar{Q}}_{\KK}\) is simply the regression of the
observed outcome on the entire history up till time \(t_{\KK-1}\)
among subjects alive and event-free at time \(t_{\KK-1}\). We further
define
\begin{align*}
    \begin{split}
      &      \bar{Q}^*_{\KK} (\bar{O}_{\KK-2},{L}_{\KK-1},
        {Y}_{\KK-1}, {D}_{\KK-1} )   = \1 \lbrace Y_{\KK-1} = D_{\KK-1}=0\rbrace \times
        \tilde{\bar{Q}}^*_{\KK} (\bar{O}_{\KK-2},{L}_{\KK-1},
        {Y}_{\KK-1}, {D}_{\KK-1} )  + \1 \lbrace Y_{\KK-1} = 1\rbrace, 
        \intertext{where,}
      &      \tilde{\bar{Q}}^*_{\KK} (\bar{O}_{\KK-2},{L}_{\KK-1},
        {Y}_{\KK-1}, {D}_{\KK-1} )  \\
      & \quad  = \sum_{a=0,1} \sum_{z=0,1}
        \tilde{\bar{Q}}_{\KK} ({A}_{\KK-1}=a,{Z}_{\KK-1}=z, {C}_{\KK-1}=0, {L}_{\KK-1}, {Y}_{\KK-1}, {D}_{\KK-1}, \bar{O}_{\KK-1})  \times    g^*_{A_{\KK-1}} (a \mid \text{Pa} (A_{\KK-1}))g^*_{Z_{\KK-1}} (z
        \mid \text{Pa} (Z_{\KK-1}))
        ,
    \end{split}
\end{align*}
obtained from \( \tilde{\bar{Q}}_{\KK} (\bar{O}_{\KK-1})\) by
integrating out the (\(\KK-1\))-specific interventional components
according to the interventional distributions defined in Section
\ref{sec:hypothetical:interventions}. Note that the `\({}^*\)'
superscript indicates that we have intervened out (here at time-point
\(\KK-1\)).

Similarly, we define, now iteratively backwards for
\(l=\KK-1, \ldots, 1\):
\begin{align*}
  & \bar{Q}_{l} (\bar{O}_{l-1})
    =
    \1 \lbrace Y_{l-1} = D_{l-1}=0\rbrace \times
    \underbrace{\EE_P  \big[  \bar{Q}^*_{l+1} (\bar{O}_{l-1}, {L}_{l}, {Y}_{l}, {D}_{l})
    \mid \bar{O}_{l-1}\big]}_{=: \tilde{\bar{Q}}_{l} (\bar{O}_{l-1})} + \,\1 \lbrace Y_{l-1} = 1\rbrace ,
    \intertext{and, }
  &  \bar{Q}^*_{l} (\bar{O}_{l-2}, {L}_{l-1}, {Y}_{l-1}, {D}_{l-1})   =
    \1 \lbrace Y_{l-1} = D_{l-1}=0\rbrace \times
    \tilde{\bar{Q}}^*_{l} (\bar{O}_{l-2}, {L}_{l-1}, {Y}_{l-1}, {D}_{l-1})
    +        \1 \lbrace Y_{l-1} = 1\rbrace,
    \intertext{where,}
  &  \tilde{\bar{Q}}^*_{l} (\bar{O}_{l-2}, {L}_{l-1}, {Y}_{l-1}, {D}_{l-1}) =\sum_{a=0,1} \sum_{z=0,1}
    \tilde{\bar{Q}}_{l} ( {A}_{l-1} = a, {Z}_{l-1}= z, C_{l-1}=0, \bar{L}_{l-1}, \bar{Y}_{l-1}, \bar{D}_{l-1}) \\[-0.2cm]
  & \qquad\qquad\qquad\qquad\qquad\qquad\qquad\qquad\qquad\qquad\qquad
    \times   Q^*_{A_{l-1}} (a \mid \text{Pa} (A_{l-1}))Q^*_{Z_{l-1}} (z
    \mid \text{Pa} (Z_{l-1}))   .
\end{align*} 
At these steps, the non-interventional components
\(L_{l}, Y_{l}, D_{l}\) are first integrated out according to the
observed data distribution, and, next, the information on the
interventional components \({A}_{l-1},{Z}_{l-1}, C_{l-1}\) are
integrated out with respect to their respective interventional
distributions. Notice that the interventions on concomitant treatment
as defined in Table \ref{table:overview:drop-in:interventions} of the
main paper ensure that exposure to concomitant treatment at each
time-point follows the same distribution among the event-free
survivors, irrespective of the past of the randomized treatment and
the time-dependent covariates.


\subsection{Targeted and non-targeted estimation}
\label{sec:targeting}

The iterative scheme laid out in Section
\ref{sec:representation:iterated:expectations} suggests a sequential
estimation procedure where information at all longitudinal time-points
are integrated out one by one, in regression steps for the
non-interventional components and according to the interventional
distributions for the interventional components. This process defines
a sequence of estimators
\begin{align}
\hat{\bar{Q}}_n^{\mathrm{gcomp}} = \big(  \hat{\bar{Q}}^{*,\mathrm{gcomp}}_{\KK},
  \hat{\bar{Q}}^{*,\mathrm{gcomp}}_{\KK-1}, \ldots,  \hat{\bar{Q}}^{*,\mathrm{gcomp}}_{2},
  \hat{\bar{Q}}^{*,\mathrm{gcomp}}_{1}\big) ,
  \label{eq:sequence:Qs}
\end{align}
where \( \hat{\bar{Q}}^{\mathrm{gcomp}}_{1}\) only depends on \(L_0\)
and defines a (non-targeted) estimator
\(\hat{\psi}_{n}^{\mathrm{gcomp}} = \frac{1}{n} \sum_{i=1}^n
\hat{\bar{Q}}^{*,\mathrm{gcomp}}_{1} (L_{0,i}) \) for the target
parameter.  Targeted estimation, on the other hand, involves an extra
step at each longitudinal time-point \(t_l\) where the estimators for
each \(\bar{Q}_{1,l}\) is updated to solve the \(l\)-specific part of
the efficient influence function. We describe this procedure below,
presenting first the efficient influence function and then the
procedure by which estimators are \(\bar{Q}_{1,l}\) updated.

\subsection{Efficient influence function}
\label{sec:eic}

Each target estimand can be represented as \(\psi = \Psi(P_0)\),
where \(\Psi \, : \, \mathcal{M}\rightarrow \R\) has efficient
influence function given by
\begin{align}
  & \phi( P) (O) = \sum_{l=1}^{\KK} \phi_l ( P) (O)
    + \bar{Q}^*_{1} ({L}_{0}) -  \EE_{P^*} [  Y_{\KK} ],
    \label{eq:efficient:ic}
\end{align}
with the \(l\)-specific components defined by
\begin{align*}
  & \phi_l ( P) (O) =
    H_l (g) (\bar{O}_{l-1})  \big(  \bar{Q}^*_{l+1}
    (\bar{O}_{l-2}, {L}_{l-1}, {Y}_{l-1}, {D}_{l-1})
    -  \bar{Q}_{l} (\bar{O}_{l-1}) \big) , 
\end{align*}
for \(l=1,\ldots, \KK\). The structure of the efficient influence
curve presented in \eqref{eq:efficient:ic} corresponds to that
presented for general longitudinal data structures in Appendix A.7 of
van der Laan and Rose 2011\cite{van2011targeted}; the only difference
is the clever weights that we defined in \eqref{eq:clever:weight}
which for our setting includes indicator functions tracking the
survival and event status of subjects.

\subsection{Targeting procedure}

The targeted estimation procedure combines the sequential regression
procedure from Section \ref{sec:representation:iterated:expectations}
with targeted update steps along parametric submodels at each
time-point to solve the efficient influence curve equation.  Here we
will assume at hand an estimator \(\hat{g}_n\) for the distributions
of all interventional components.  For each \(l\), one starts with an
estimator \(\hat{\bar{Q}}^{*,\mathrm{tmle}}_{l+1}\) for
\({\bar{Q}}^*_{l+1}\) and an estimator \(\hat{\bar{Q}}_{l}\) for
\({\bar{Q}}_{l}\) obtained by a regression of
\(\hat{\bar{Q}}^{*,\mathrm{tmle}}_{l+1}\) on the observed history
\(\bar{O}_{l-1}\). Next, a targeted intercept-only parametric
regression step is carried out with
\(\hat{\bar{Q}}^{*,\mathrm{tmle}}_{l+1}\) as outcome and with weights
given by estimated clever weights
\( H_l (\hat{g}_n) (\bar{O}_{l-1})\). Denoting the estimate of the
intercept by \(\hat{\eps}_{n,l}\), the estimator \(\hat{\bar{Q}}_{l}\)
is updated as follows
\begin{align*}
  \hat{\bar{Q}}^{\mathrm{tmle}}_{l} (\bar{O}_{l-1}) = \mathrm{expit} \big(\mathrm{logit}
    \hat{\bar{Q}}_{l} ( \bar{O}_{l-1}) + \hat{\eps}_{n,l}\big) ,
\end{align*}
and, per construction of the targeted regression step, we then have
that
\begin{align*}
\frac{1}{n}\sum_{i=1}^n H_l (\hat{g}_n) (\bar{O}_{l-1, i})  \big(  \hat{\bar{Q}}^{*,\mathrm{tmle}}_{l+1}
    (\bar{O}_{l-2,i}, {L}_{l-1,i}, {Y}_{l-1,i}, {D}_{l-1,i})
    -  \hat{\bar{Q}}^{\mathrm{tmle}}_{l} (\bar{O}_{l-1,i}) \big) = 0,
\end{align*}
i.e., the pair
\(
\hat{\bar{Q}}^{*,\mathrm{tmle}}_{l+1},\hat{\bar{Q}}^{\mathrm{tmle}}_{l}\)
solves the \(l\)-specific contribution to the efficient influence
curve equation.  Repeating these steps iteratively for
\(l= \KK, \ldots, 1\), obtaining \(\hat{\bar{Q}}^{\mathrm{tmle}}_{l}\)
from \( \hat{\bar{Q}}^{*,\mathrm{tmle}}_{l+1}\) in a targeted
regression, and next obtaining
\( \hat{\bar{Q}}^{*,\mathrm{tmle}}_{l}\) from
\(\hat{\bar{Q}}^{\mathrm{tmle}}_{l}\) by integrating out
interventional components according to the interventional
distributions, one obtains a targeted sequence of estimators
\begin{align}
\hat{Q}_n^{\mathrm{tmle}} = \big(  \hat{\bar{Q}}^{\mathrm{tmle}}_{\KK}, \hat{\bar{Q}}^{*,\mathrm{tmle}}_{\KK}, \hat{\bar{Q}}^{\mathrm{tmle}}_{\KK-1},
  \hat{\bar{Q}}^{*,\mathrm{tmle}}_{\KK-1}, \ldots, \hat{\bar{Q}}^{\mathrm{tmle}}_{2}, \hat{\bar{Q}}^{*,\mathrm{tmle}}_{2},
  \hat{\bar{Q}}^{\mathrm{tmle}}_{1}, \hat{\bar{Q}}^{*,\mathrm{tmle}}_{1}\big) ,
  \label{eq:sequence:Qs}
\end{align}
where \( \hat{\bar{Q}}^{*,\mathrm{tmle}}_{1}\) defines a targeted
estimator
\(\hat{\psi}_{n}^{\mathrm{tmle}} = \frac{1}{n} \sum_{i=1}^n
\hat{\bar{Q}}^{*,\mathrm{tmle}}_{1} (L_{0,i}) \) for the target
parameter. Particularly, the sequence of estimators
\eqref{eq:sequence:Qs} solves the efficient influence curve equation,
i.e., \(\frac{1}{n}\sum_{i=1}^n \phi(\hat{P}^{\mathrm{tmle}}_n) = 0 \)
where
\( \hat{P}^{\mathrm{tmle}}_n= ( \hat{Q}^{\mathrm{tmle}}_n,
\hat{g}_n)\).  {Under further conditions
  \cite{van2006targeted,van2017generally,van2011targeted}
  related to
  the estimation of \(Q\) and \(g\) (the product of their rates of
  convergence have to be faster than \(n^{-1/2}\)),} this again
implies that,
\begin{align*}
  \sqrt{n} \big(  \hat{\psi}^{\mathrm{tmle}}_n - \psi_0 \big) = \sqrt{n} \,
  \mathbb{P}_n \phi(P_0) + o_P(1), 
\end{align*}
and we can use the asymptotic normal distribution
\begin{align*}
  \sqrt{n}\, \big( \hat{\psi}^{\mathrm{tmle}}_n
  - \psi_0 \big)
  \overset{\mathcal{D}}{\rightarrow} \mathcal{N} (0, {P}_0  \phi ( P_0
  )^2),
\end{align*}
to provide an approximate confidence interval. Specifically, the
asymptotic variance of the estimator is equal to the variance of the
efficient influence function and can be estimated by
\(\hat{\sigma}_n^2 /n\) where
\(\hat{\sigma}_n^2 = \mathbb{P}_n (\phi (
\hat{P}^{\mathrm{tmle}}_n))^2\).



\begin{table}[!ht]
  \centering
\begin{tabular}{rlll}
  \\[-0.25cm]
\toprule \\[-0.15cm]
\multirow{4}{*}{\(  \bar{Q}_{l} (\bar{O}_{l-1})\)} 
& obtained from \( \bar{Q}^*_{l+1} \)  &  \\[0.1cm]
& by   marginalization    & \( = \EE_{P_{q,g^*}}[ Y_{\KK}  \mid \bar{O}_{l-1} ]\) \\[0.1cm]
& over  non-interventional  & \\[0.1cm] 
  & nodes \(L_{l}, Y_{l}, D_{l}\) &   \\[0.4cm]
 \multirow{4}{*}{ \(    \bar{Q}^*_{l} (\bar{O}_{l-2}, {L}_{l-1}, {Y}_{l-1}, {D}_{l-1}) \)}
&
 obtained from \( \bar{Q}_{l} \) by &
               \\[0.1cm]
   &  marginalization  over 
                                          &    \( = \EE_{P_{q,g^*}}[ Y_{\KK}  \mid \bar{O}_{l-2}, {L}_{l-1}, {Y}_{l-1}, {D}_{l-1} ]\)
               \\[0.1cm]
&   interventional  nodes &  \\[0.1cm]
&  \(C_{l-1}, A_{l-1}, Z_{l-1}\) & \\[0.4cm]
  \hline\\[-0.2em]
\end{tabular}
  \caption{Overview of notation used for the sequential regression
    procedure.}
\label{tab:overview:Q:notation}
\end{table}

\section{Causal interpretation of the target parameter}

{It is general result that the g-computation formula identifies the
  post-interventional distribution under the stated causal assumptions
  Section \ref{sec:identifiability}
  \cite{robins1986new,gill2001causal}. For our purposes, this mean
  particularly that
  \(\psi = \EE[ \bar{Q}^*_{1} (L_0) ] = \EE[ Y^*_{\KK} ] \), where
  \(Y^*_{k_0}\) is the counterfactual outcome we would have observed
  for a subject had they been assigned to intervention \(g^*\). Here
  we repeat this proof, particularly how the identifiability
  assumptions of Section \ref{sec:identifiability} allow the causal
  interpretation \eqref{eq:causal:estimand}, for completeness.}

Particularly, we show that it holds that
\begin{align*}
  \bar{Q}_{l} (\bar{O}_{l-1})
  & \overset{a}{=}
    \EE_{P_{q,g^*}}[ Y_{\KK}  \mid \bar{O}_{l-1} ]
    \overset{b}{=} \EE[ Y^*_{\KK}  \mid \bar{O}_{l-1} ]
  \\
  \bar{Q}^*_{l} (\bar{O}_{l-2}, {L}_{l-1}, {Y}_{l-1}, {D}_{l-1})
  & \overset{a}{=}  \EE_{P_{q,g^*}}[ Y_{\KK}  \mid \bar{O}_{l-2}, {L}_{l-1}, {Y}_{l-1}, {D}_{l-1} ]
    \overset{b}{=} \EE[ Y^*_{\KK}   \mid \bar{O}_{l-2}, {L}_{l-1}, {Y}_{l-1}, {D}_{l-1} ], 
\end{align*}
where the equalities marked by `\(b\)' requires the causal assumptions
from Section 3.3 (and \(Y^*_{k_0}\) is the counterfactual outcome we
would have observed for a subject had they been assigned to
intervention \(g^*\)). Now, first, the equalities marked by `\(a\)'
follow by repeated use of Fubini's theorem. One starts by noting that
\begin{align*}
  \bar{Q}_{\KK} (\bar{O}_{\KK-1})
  &=  \EE_P  [Y_{\KK} \mid \bar{A}_{\KK-1},\bar{Z}_{\KK-1}, \bar{L}_{\KK-1},
    \bar{Y}_{\KK-1}, \bar{D}_{\KK-1},  \bar{C}_{\KK-1}] \\
  &=  \EE_{P_{q, g^*}}  [Y_{\KK} \mid \bar{A}_{\KK-1},\bar{Z}_{\KK-1}, \bar{L}_{\KK-1},
    \bar{Y}_{\KK-1}, \bar{D}_{\KK-1},  \bar{C}_{\KK-1}]
\end{align*}
and that
\begin{align*}
  \bar{Q}^*_{\KK} (\bar{O}_{\KK-2},{L}_{\KK-1},
  {Y}_{\KK-1}, {D}_{\KK-1} )
  &=  \EE_{P_{q, g^*}}  [ \bar{Q}_{\KK} (\bar{O}_{\KK-1}) \mid \bar{O}_{\KK-2},{L}_{\KK-1},
    {Y}_{\KK-1}, {D}_{\KK-1} ] \\
  &=  \EE_{P_{q, g^*}}  [ Y_{\KK} \mid \bar{O}_{\KK-2},{L}_{\KK-1},
    {Y}_{\KK-1}, {D}_{\KK-1} ] . 
\end{align*}
Next, the same observations are made iteratively backwards for
\(l=\KK-1, \ldots, 1\), assuming that  `\(a\)' holds for \(l+1\) so that 
\begin{align*}
  \bar{Q}_{l} (\bar{O}_{l-1})
  &=  \EE_P  [\bar{Q}^*_{l+1} (\bar{O}_{l-1},{L}_{l},
    {Y}_{l}, {D}_{l} )\mid \bar{A}_{l-1},\bar{Z}_{l-1}, \bar{L}_{l-1},
    \bar{Y}_{l-1}, \bar{D}_{l-1},  \bar{C}_{l-1}] \\
  &=  \EE_{P_{q, g^*}}  [\bar{Q}^*_{l+1} (\bar{O}_{l-1},{L}_{l},
    {Y}_{l}, {D}_{l} )\mid \bar{A}_{l-1},\bar{Z}_{l-1}, \bar{L}_{l-1},
    \bar{Y}_{l-1}, \bar{D}_{l-1},  \bar{C}_{l-1}] \\
  &=  \EE_{P_{q, g^*}}  [Y_{l} \mid \bar{A}_{l-1},\bar{Z}_{l-1}, \bar{L}_{l-1},
    \bar{Y}_{l-1}, \bar{D}_{l-1},  \bar{C}_{l-1}]
\end{align*}
and that
\begin{align*}
  \bar{Q}^*_{l} (\bar{O}_{l-2},{L}_{l-1},
  {Y}_{l-1}, {D}_{l-1} )
  &=  \EE_{P_{q, g^*}}  [ \bar{Q}_{l} (\bar{O}_{l-1}) \mid \bar{O}_{l-2},{L}_{l-1},
    {Y}_{l-1}, {D}_{l-1} ] \\
  &=  \EE_{P_{q, g^*}}  [ Y_{l} \mid \bar{O}_{l-2},{L}_{l-1},
    {Y}_{l-1}, {D}_{l-1} ] . 
\end{align*}
At \(l=1\) one then obtains that
\(\bar{Q}^*_{1} (L_0) =\EE_{P_{q,g^*}}[ Y_{\KK} \mid L_0 ]\) and the
target parameter is identified by marginalizing the latter over the
distribution of \(L_0\), i.e.,
\(\psi = \EE[ \bar{Q}^*_{1} (L_0) ]\).

To verify the causal interpretation, note that the identification
process outlined above can also be represented as
\begin{align*}
  \EE [ \EE_{P_{q, g^*}}[ \EE[ \cdots \EE[ \EE_{P_{q, g^*}}[ \EE[ Y_{\KK}
  \mid \bar{O}_{\KK-1}]  \mid \bar{O}_{\KK-2},{L}_{\KK-1},
  {Y}_{\KK-1}, {D}_{\KK-1}] \mid \bar{O}_{\KK-2}] \cdots \mid A_0, Z_0, L_0] \mid L_0] ]. 
\end{align*}
From here, the innermost expectation
\(\EE[ Y_{\KK} \mid \bar{O}_{\KK-1}]\) is first replaced by
\(\EE[ Y^*_{\KK} \mid \bar{O}_{\KK-1}]\)
. Then one proceeds iteratively backwards for
\(l=\KK-1, \ldots, 1\), assuming that `\(b\)' holds for \(l+1\) so
that
\begin{align*}
  \bar{Q}^*_{l+1} (\bar{O}_{l-1}, {L}_{l}, {Y}_{l}, {D}_{l})
  = \EE[ Y^*_{\KK}   \mid \bar{O}_{l-1}, {L}_{l}, {Y}_{l}, {D}_{l} ].  
\end{align*}
From here note that 
\begin{align*}
  \bar{Q}_{l} (\bar{O}_{l-1})
  =  \EE[  \bar{Q}^*_{l+1} (\bar{O}_{l-1}, {L}_{l}, {Y}_{l}, {D}_{l})   \mid \bar{O}_{l-1} ]
  = \EE[ Y^*_{\KK}   \mid \bar{O}_{l-1} ] ,   
\end{align*}
and then, by no unmeasured confounding, that
\begin{align*}
  \EE[ Y^*_{\KK} \mid \bar{O}_{l-1}] = 
    \EE[ Y^*_{\KK} \mid \bar{O}_{l-2},{L}_{l-1},
  {Y}_{l-1}, {D}_{l-1}] =   \bar{Q}^*_{l} (\bar{O}_{l-2},{L}_{l-1},
  {Y}_{l-1}, {D}_{l-1} ).
\end{align*}
One achieves then that
\(\bar{Q}^*_{1} ({L}_{0}) = \EE[ Y^*_{\KK} \mid {L}_{0} ]\), and
marginalization over \(L_0\) according to the observed data
distribution identifies \(\EE[ Y^*_{\KK} ]\).


\section*{Bibliography}

\begin{thebibliography}{31}
\providecommand{\natexlab}[1]{#1}
\providecommand{\url}[1]{\texttt{#1}}
\expandafter\ifx\csname urlstyle\endcsname\relax
  \providecommand{\doi}[1]{doi: #1}\else
  \providecommand{\doi}{doi: \begingroup \urlstyle{rm}\Url}\fi

\bibitem[Bang and Robins(2005)]{bang2005doubly}
H.~Bang and J.~M. Robins.
\newblock Doubly robust estimation in missing data and causal inference models.
\newblock \emph{Biometrics}, 61\penalty0 (4):\penalty0 962--973, 2005.

\bibitem[Bethel et~al.(2020)Bethel, Stevens, Buse, Choi, Gustavson, Iqbal,
  Lokhnygina, Mentz, Patel, and {\"O}hman]{bethel2020exploring}
M.~A. Bethel, S.~R. Stevens, J.~B. Buse, J.~Choi, S.~M. Gustavson, N.~Iqbal,
  Y.~Lokhnygina, R.~J. Mentz, R.~A. Patel, and P.~{\"O}hman.
\newblock Exploring the possible impact of unbalanced open-label drop-in of
  glucose-lowering medications on exscel outcomes.
\newblock \emph{Circulation}, 141\penalty0 (17):\penalty0 1360--1370, 2020.

\bibitem[Didelez et~al.(2006)Didelez, Dawid, and Geneletti]{didelez2006direct}
V.~Didelez, P.~Dawid, and S.~Geneletti.
\newblock Direct and indirect effects of sequential treatments.
\newblock \emph{Proceedings of the 22nd Annual Conference on Uncertainty in
  Artifical Intelligence}, pages 138--146, 2006.

\bibitem[Gill and Robins(2001)]{gill2001causal}
R.~D. Gill and J.~M. Robins.
\newblock Causal inference for complex longitudinal data: the continuous case.
\newblock \emph{Annals of Statistics}, pages 1785--1811, 2001.

\bibitem[Hern{\'a}n(2010)]{hernan2010hazards}
M.~A. Hern{\'a}n.
\newblock The hazards of hazard ratios.
\newblock \emph{Epidemiology (Cambridge, Mass.)}, 21\penalty0 (1):\penalty0 13,
  2010.

\bibitem[Hernan and Robins(2020)]{hernan2020causal}
M.~A. Hernan and J.~Robins.
\newblock Causal inference: What if. boca raton: Chapman \& hill/crc.
\newblock 2020.

\bibitem[Hern{\'a}n and Robins(2016)]{hernan2016using}
M.~A. Hern{\'a}n and J.~M. Robins.
\newblock Using big data to emulate a target trial when a randomized trial is
  not available.
\newblock \emph{American journal of epidemiology}, 183\penalty0 (8):\penalty0
  758--764, 2016.

\bibitem[ICH(2019)]{international2019addendum}
ICH.
\newblock Addendum on estimands and sensitivity analysis in clinical trials to
  the guideline on statistical principles for clinical trials {E}9 ({R}1).
\newblock \emph{Fed Regist}, pages 1--19, 2019.

\bibitem[Lendle et~al.(2017)Lendle, Schwab, Petersen, and {{v}an {d}er
  Laan}]{ltmleRpackage}
S.~D. Lendle, J.~Schwab, M.~L. Petersen, and M.~J. {{v}an {d}er Laan}.
\newblock {ltmle}: An {R} package implementing targeted minimum loss-based
  estimation for longitudinal data.
\newblock \emph{Journal of Statistical Software}, 81\penalty0 (1):\penalty0
  1--21, 2017.
\newblock \doi{10.18637/jss.v081.i01}.

\bibitem[Marso et~al.(2016)Marso, Daniels, Brown-Frandsen, Kristensen, Mann,
  Nauck, Nissen, Pocock, Poulter, and Ravn]{marso2016liraglutide}
S.~P. Marso, G.~H. Daniels, K.~Brown-Frandsen, P.~Kristensen, J.~F.~E. Mann,
  M.~A. Nauck, S.~E. Nissen, S.~Pocock, N.~R. Poulter, and S.~W. M. S. M. Z. B.
  B. R. M. B. J.~B. Ravn, L.~S.
\newblock Liraglutide and cardiovascular outcomes in type 2 diabetes.
\newblock \emph{New England Journal of Medicine}, 375\penalty0 (4):\penalty0
  311--322, 2016.

\bibitem[Martinussen et~al.(2020)Martinussen, Vansteelandt, and
  Andersen]{martinussen2020subtleties}
T.~Martinussen, S.~Vansteelandt, and P.~K. Andersen.
\newblock Subtleties in the interpretation of hazard contrasts.
\newblock \emph{Lifetime Data Analysis}, 26:\penalty0 833--855, 2020.

\bibitem[McGuire et~al.(2022)McGuire, D’Alessio, Nicholls, Nissen, Riesmeyer,
  Pavo, Sethuraman, Heilmann, Kaiser, and Weerakkody]{mcguire2022transitioning}
D.~K. McGuire, D.~D’Alessio, S.~J. Nicholls, S.~E. Nissen, J.~S. Riesmeyer,
  I.~Pavo, S.~Sethuraman, C.~R. Heilmann, J.~J. Kaiser, and G.~J. Weerakkody.
\newblock Transitioning to active-controlled trials to evaluate cardiovascular
  safety and efficacy of medications for type 2 diabetes.
\newblock \emph{Cardiovascular Diabetology}, 21\penalty0 (1):\penalty0 1--9,
  2022.

\bibitem[Michiels et~al.(2021)Michiels, Sotto, Vandebosch, and
  Vansteelandt]{michiels2021novel}
H.~Michiels, C.~Sotto, A.~Vandebosch, and S.~Vansteelandt.
\newblock A novel estimand to adjust for rescue treatment in randomized
  clinical trials.
\newblock \emph{Statistics in Medicine}, 40\penalty0 (9):\penalty0 2257--2271,
  2021.

\bibitem[Pearl(2001)]{pearl2001direct}
J.~Pearl.
\newblock Direct and indirect effects.
\newblock \emph{Proceedings of the seventeenth conference on uncertainty in
  artificial intelligence}, pages 411--420, 2001.

\bibitem[Petersen et~al.(2014)Petersen, Schwab, Gruber, Blaser, Schomaker, and
  {v}an~{d}er Laan]{petersen2014targeted}
M.~Petersen, J.~Schwab, S.~Gruber, N.~Blaser, M.~Schomaker, and M.~{v}an~{d}er
  Laan.
\newblock Targeted maximum likelihood estimation for dynamic and static
  longitudinal marginal structural working models.
\newblock \emph{Journal of causal inference}, 2\penalty0 (2):\penalty0
  147--185, 2014.

\bibitem[Petersen and van~der Laan(2014)]{petersen2014causal}
M.~L. Petersen and M.~J. van~der Laan.
\newblock Causal models and learning from data: integrating causal modeling and
  statistical estimation.
\newblock \emph{Epidemiology (Cambridge, Mass.)}, 25\penalty0 (3):\penalty0
  418, 2014.

\bibitem[Petersen et~al.(2012)Petersen, Porter, Gruber, Wang, and van~der
  Laan]{petersen2012diagnosing}
M.~L. Petersen, K.~E. Porter, S.~Gruber, Y.~Wang, and M.~J. van~der Laan.
\newblock Diagnosing and responding to violations in the positivity assumption.
\newblock \emph{Statistical methods in medical research}, 21\penalty0
  (1):\penalty0 31--54, 2012.

\bibitem[Polley et~al.(2011)Polley, Rose, and van~der Laan]{polley2011super}
E.~C. Polley, S.~Rose, and M.~J. van~der Laan.
\newblock Super learning.
\newblock In \emph{Targeted Learning}, pages 43--66. Springer, 2011.

\bibitem[Robins(1986)]{robins1986new}
J.~Robins.
\newblock A new approach to causal inference in mortality studies with a
  sustained exposure period—application to control of the healthy worker
  survivor effect.
\newblock \emph{Mathematical modelling}, 7\penalty0 (9-12):\penalty0
  1393--1512, 1986.

\bibitem[Robins and Greenland(1992)]{robins1992identifiability}
J.~M. Robins and S.~Greenland.
\newblock Identifiability and exchangeability for direct and indirect effects.
\newblock \emph{Epidemiology}, pages 143--155, 1992.

\bibitem[Stitelman et~al.(2011)Stitelman, De~Gruttola, Wester, and van~der
  Laan]{stitelman2011rcts}
O.~M. Stitelman, V.~De~Gruttola, C.~W. Wester, and M.~J. van~der Laan.
\newblock Rcts with time-to-event outcomes and effect modification parameters.
\newblock In \emph{Targeted Learning}, pages 271--298. Springer, 2011.

\bibitem[{v}an~{d}er Laan(2017)]{van2017generally}
M.~J. {v}an~{d}er Laan.
\newblock A generally efficient targeted minimum loss based estimator based on
  the highly adaptive lasso.
\newblock \emph{The International Journal of Biostatistics}, 13\penalty0 (2),
  2017.

\bibitem[{v}an~{d}er Laan and Gruber(2012)]{van2012targeted}
M.~J. {v}an~{d}er Laan and S.~Gruber.
\newblock Targeted minimum loss based estimation of causal effects of multiple
  time point interventions.
\newblock \emph{The international journal of biostatistics}, 8\penalty0 (1),
  2012.

\bibitem[van~der Laan and Petersen(2008)]{van2008direct}
M.~J. van~der Laan and M.~L. Petersen.
\newblock Direct effect models.
\newblock \emph{The international journal of biostatistics}, 4\penalty0 (1),
  2008.

\bibitem[{v}an~{d}er Laan and Rose(2011)]{van2011targeted}
M.~J. {v}an~{d}er Laan and S.~Rose.
\newblock \emph{Targeted learning: causal inference for observational and
  experimental data}.
\newblock Springer Science \& Business Media, 2011.

\bibitem[{v}an~{d}er Laan and Rose(2018)]{van2018targeted}
M.~J. {v}an~{d}er Laan and S.~Rose.
\newblock \emph{Targeted learning in data science: causal inference for complex
  longitudinal studies}.
\newblock Springer, 2018.

\bibitem[{v}an~{d}er Laan and Rubin(2006)]{van2006targeted}
M.~J. {v}an~{d}er Laan and D.~Rubin.
\newblock Targeted maximum likelihood learning.
\newblock \emph{The International Journal of Biostatistics}, 2\penalty0 (1),
  2006.

\bibitem[{v}an~der Laan et~al.(2007){v}an~der Laan, Polley, and
  Hubbard]{van2007super}
M.~J. {v}an~der Laan, E.~C. Polley, and A.~E. Hubbard.
\newblock Super learner.
\newblock \emph{Statistical applications in genetics and molecular biology},
  6\penalty0 (1), 2007.

\bibitem[VanderWeele and Tchetgen~Tchetgen(2017)]{vanderweele2017mediation}
T.~J. VanderWeele and E.~J. Tchetgen~Tchetgen.
\newblock Mediation analysis with time varying exposures and mediators.
\newblock \emph{Journal of the Royal Statistical Society Series B: Statistical
  Methodology}, 79\penalty0 (3):\penalty0 917--938, 2017.

\bibitem[Vansteelandt et~al.(2019)Vansteelandt, Linder, Vandenberghe, Steen,
  and Madsen]{vansteelandt2019mediation}
S.~Vansteelandt, M.~Linder, S.~Vandenberghe, J.~Steen, and J.~Madsen.
\newblock Mediation analysis of time-to-event endpoints accounting for
  repeatedly measured mediators subject to time-varying confounding.
\newblock \emph{Statistics in medicine}, 38\penalty0 (24):\penalty0 4828--4840,
  2019.

\bibitem[Zheng and van~der Laan(2017)]{zheng2017longitudinal}
W.~Zheng and M.~van~der Laan.
\newblock Longitudinal mediation analysis with time-varying mediators and
  exposures, with application to survival outcomes.
\newblock \emph{Journal of causal inference}, 5\penalty0 (2), 2017.

\end{thebibliography}
\end{document}